\documentclass{article}
\usepackage[english]{babel}
\usepackage[utf8]{inputenc}
\usepackage{amsmath}
\usepackage{amsfonts}
\usepackage{amssymb}
\usepackage{graphicx}
\usepackage[style=apa,citestyle=apa, backend=biber, uniquename=false, mincitenames=1, maxcitenames=2, citetracker]{biblatex}
\usepackage{setspace}
\usepackage{wrapfig}
\usepackage{fancyhdr}
\usepackage{tabularx}
\usepackage[a4paper,margin=3.5cm]{geometry}
\usepackage[right]{eurosym}
\usepackage[printonlyused]{acronym}
\usepackage{subcaption}
\usepackage{floatflt}
\usepackage[usenames,dvipsnames]{color}
\usepackage{colortbl}
\usepackage{paralist}
\usepackage{array}
\usepackage{indentfirst}
\usepackage{titlesec}
\usepackage{longtable}
\usepackage[colorlinks=true,linkcolor=black,urlcolor=black,citecolor=black,pdfpagelabels=true]{hyperref}
\usepackage{tablefootnote}
\usepackage{csquotes}
\usepackage{mathtools}
\usepackage{booktabs,caption}
\usepackage[flushleft]{threeparttable}
\usepackage{hyperref}
\usepackage[bottom,flushmargin]{footmisc}
\usepackage{multicol}
\usepackage{pdflscape}
\usepackage{titlesec}
\usepackage{titling}
\usepackage{authblk}
\usepackage[inline]{enumitem}
\usepackage{xcolor, soul}
\sethlcolor{white}
\setulcolor{white}
\setstcolor{white}
\usepackage{pifont}
\usepackage{threeparttable}
\usepackage{threeparttablex}
\usepackage{float}
\usepackage{comment}
\usepackage{multirow}
\usepackage{chngpage}
\usepackage{everypage}
\usepackage{lipsum}

\newcommand{\Lpagenumber}{\ifdim\textwidth=\linewidth\else\bgroup
  \dimendef\margin=0 
  \ifodd\value{page}\margin=\oddsidemargin
  \else\margin=\evensidemargin
  \fi
  \raisebox{\dimexpr -\topmargin-\headheight-\headsep-0.5\linewidth}[0pt][0pt]{%
    \rlap{\hspace{\dimexpr \margin+\textheight+\footskip}%
    \llap{\rotatebox{90}{\thepage}}}}%
\egroup\fi}
\AddEverypageHook{\Lpagenumber}%

\makeatletter
\AtEveryCitekey{%
  \ifciteseen
    {}
    {\defcounter{maxnames}{6}}%
}
\makeatother

\def\plusplus{{\nolinebreak[4]\hspace{-.05em}\raisebox{.4ex}{\tiny\bf ++}}}

\setul{0.3ex}{.5ex}

\newcolumntype{H}{>{\setbox0=\hbox\bgroup}c<{\egroup}@{}}

\providecommand{\keywords}[1]
{
  \noindent
  \small	
  \textbf{Keywords:} #1
}

\providecommand{\jelclass}[1]
{
  \noindent
  \small	
  \textbf{JEL classification:} #1
}

\numberwithin{equation}{section}

\title{\huge The Effect of Sport in Online Dating:\\Evidence from Causal Machine Learning\thanks{\noindent This research project is part of the National Research Programme 'Big Data' (NRP 75) of the Swiss National Science Foundation (SNSF). Further information on the National Research Programme can be found at \url{www.nrp75.ch} or \url{https://bigdata-dialog.ch}. This work has been co-funded by the GFF-IPF Grant of the Basic Research Fund of the University of St.Gallen. We gratefully acknowledge the data access provided by Parship.de (PE Digital GmbH, Hamburg). A previous version of this paper was presented at research seminars of the University of St.Gallen, the GESIS Spring Seminar in Cologne and the Causal Machine Learning Workshop in St.Gallen. We thank participants, in particular Daniel Goller, Fabian Muny and Anthony Strittmatter for helpful comments. The usual disclaimer applies.}\\
~\\
}

\author{Daniel Boller\thanks{\noindent Daniel Boller is also affiliated with the Asian Development Bank. Email: \url{daniel.boller@unisg.ch}}}
\author{Michael Lechner\thanks{\noindent Michael Lechner is also affiliated with CEPR, London, CESifo, Munich, IAB, Nuremberg, and IZA, Bonn.\\
Email: \url{michael.lechner@unisg.ch}} }
\author{Gabriel Okasa\thanks{\noindent Email: \url{gabriel.okasa@unisg.ch}}}

\affil{SEW-HSG\\ Swiss Institute for Empirical Economic Research\\ University of St.Gallen, Switzerland}

\begin{document}

\maketitle
\vspace{1cm}
\begin{abstract}
Online dating emerged as a key platform for human mating. Previous research focused on socio-demographic characteristics to explain human mating in online dating environments, neglecting the commonly recognized relevance of sport. This research investigates the effect of sport activity on human mating by exploiting a unique data set from an online dating platform. Thereby, we leverage recent advances in the causal machine learning literature to estimate the causal effect of sport frequency on the contact chances. We find that for male users, doing sport on a weekly basis increases the probability to receive a first message from a woman by 50\%, relatively to not doing sport at all. For female users, we do not find evidence for such an effect. In addition, for male users the effect increases with higher income.\\

\\

\jelclass{J12, Z29, C21, C45.}
\end{abstract}
\vfill

\definecolor{darkgreen}{cmyk}{1,0,0.93,0.5}
\definecolor{lightgreen}{cmyk}{0.8,0,0.95,0.25}
\definecolor{lightlightgreen}{cmyk}{0.2,0,0.235,0.065}
\definecolor{femalegreen}{rgb}{0.91, 0.45, 0.32}
\definecolor{malegreen}{rgb}{0.0, 0.55, 0.55}

\thispagestyle{empty}

\pagebreak
\onehalfspacing
\setcounter{page}{1}

\section{Introduction}

Human interactions that have traditionally taken place in physical reality have increasingly shifted to the online world and the Covid-19 pandemic has substantially accelerated this trend. Human mating is also affected by this development, resulting in numerous novel formats of online dating. Indeed, online dating emerged as pivotal instrument for human mating. \textcite{rosenfeld2019}, for instance, showed, that online dating represents the most common way for heterosexual couples to meet in the US. \textcite{cacioppo2013} furthermore showed, that more than one-third of marriages in the US (2005-2012) are attributed to an initial contact via online dating. 

Understanding the mechanisms that explain human mating in online dating environments is, in turn, decisive to elucidate the structure of societal evolution and to derive algorithms increasing the efficiency of the matching of potential partners. Explaining human mating in online dating environments relies essentially on the information that users share online, including socio-demographic, psychological, and physical traits. Indeed, previous research referred to socio-demographic \parencite[e.g., age;][]{hitsch2010a} and psychological \parencite[e.g., extroversion;][]{cuperman2009} traits to explain human mating in online dating environments \parencite[for a detailed review, see][]{eastwick2014}. Research considering physical traits, commonly interpreted as sport activity \parencite{schulte2012}, to explain human mating in online dating environments, however, remains sparse even though few research provides indications that sport activity has substantial effects on human mating \parencite{schulte2012}. However, the effect of sport activity on human mating has not yet been fully understood. This paper attempts to fill this gap. In particular, this paper is, to the best of our knowledge, the first to investigate the causal effect of sport activity on human mating in online dating environments. It is also the first paper to analyze the heterogeneity of this causal effect using the novel causal machine learning methods.

Following this notion, we leverage unique data of more than 16'000 users, forming altogether almost 180'000 interactions. The data allows us not only to map interactions among users on a second-by-second basis, including visiting a user profile and contacting a user via private message, but also to observe more than 600 user characteristics describing the socio-demographic, psychological, and importantly, physical traits, including the frequency of the sport activity. This setting allows us to create a credible research design that eliminates potential sources of endogeneity by focusing on the first, one-way interactions between users, and by observing essentially the very same information, and even beyond,  as an actual user. Hence, we can reliably identify the effect of sport activity on contact chances by relying on the conditional independence, i.e. the unconfoundedness research design. Moreover, we exploit recent advances in causal machine learning to estimate the causal effect of sport activity on contact chances in our large-dimensional setting in a very flexible way, while considering potential effect heterogeneities. In particular, we apply the Modified Causal Forest \parencite{lechner2018}, an estimator that reams the concept of Causal Trees and Forests, by allowing for multiple treatments, as applicable to our measure of sport activity. Furthermore, the Modified Causal Forest improves the splitting rule to account for selection bias and the mean correlated error. Additionally, it allows for estimation and inference on different aggregation levels in one estimation step. All of these aspects are crucial and beneficial for our research. Specifically, we can relax on the functional form assumptions, unlike classical parametric approaches, which is particularly important in large-dimensional settings as ours. Moreover, we can go beyond average effects and can flexibly investigate effect heterogeneities on various aggregation levels.

Leveraging the benefits of the Modified Causal Forest, we find different patterns for males and females. Particularly, for male users, we observe uniformly increasing contact chances by a potential female partner, for increasing levels of sport activity. Specifically, the contact chances increase by more than $50\%$ if male users practice sport on a weekly basis, relative to no sport at all. However, for female users, we do not find evidence for such an effect. Beyond the average effects, we uncover interesting effect heterogeneities both for males and females. In particular, for male users, we find that the effect of sport frequency on contact chances increases with higher income. This holds true for the income levels of the male users themselves, as well as for the income levels of the potential female partners. This implies that higher income male users enjoy a higher effect of a weekly sport activity, and that higher income female users value the regular sport activity of the potential male partners more. These heterogeneous effects are both statistically precise, as well as substantially relevant. In addition, for female users, we find indications that the effect of sport activity on contact chances increases with a higher sport frequency of the potential male partner. Furthermore, analysing the individualized effects provides additional descriptive evidence for these heterogeneous effects. It reveals further insights for potential heterogeneity mechanisms driven by education level or relationship preferences, among others. Lastly, a placebo test shows the robustness of our results.

This study contributes to research and practice as well as to the society. First, this paper provides new insights for the literature on human mating by demonstrating that sport activity, a key behavioral trait, affects human mating. Second, this paper supports social science research in assessing causal effects in large-dimensional data environments by showcasing an empirical approach, which allows for a very flexible estimation of average effects as well as a systematic assessment of underlying heterogeneities. Third, this paper helps individuals to increase their dating success by exhibiting how sport activities can contribute to the likelihood to be recognized by potential partners, finally highlighting the relevance of sport activity not only from a health but also from a human mating perspective. Finally, this paper serves product developers to improve the architecture of online dating platforms by highlighting the relevance of sport activity, while considering effect heterogeneities (e.g., demographic characteristics) at the same time.

This paper is structured as follows. Section \ref{sec:literature} provides a short overview on prior work related to our research. Section \ref{sec:setup} describes the online dating platform and the respective data. Section \ref{sec:empirics} explains the empirical approach, including the identification strategy and the estimation method. Following this, Section \ref{sec:results} presents the results, comprising the average and disaggregated effects. Section \ref{sec:discussion} discusses the results and the implications for research, practice, and society. 

\section{Literature}\label{sec:literature}

In this section we briefly describe prior work related to our research, comprising literature on sport activity in general as well as literature on sport activity and human mating. 

\subsection{Sport Activity}

Sport activity has been ascribed relevant effects on human life, including physical and mental health as well as social outcomes, some of which are summarized next. 

First, sport activity was shown to affect health outcomes. For instance \textcite{warburton2006} confirmed, based on an extensive review of the literature, that sport activity facilitates the prevention of several chronic diseases (e.g., cardiovascular disease and diabetes). In a similar vein, \textcite{humphreys2014} found, that sport activity reduces self-reported incidences of diabetes, high blood pressure, heart disease, asthma, and arthritis \parencites()()[for a review, see][and]{penedo2005}{eime2013}.

Second, sport activity was demonstrated to enfold effects on mental health. \textcite{hillman2008}, for instance, showed that sport activity enhances cognition and brain functions \parencites()()[for a review, see][and]{strong2005}{janssen2010}. Moreover, sport activity was shown to increase self-reported life satisfaction and happiness \parencites()(){huang2012}[]{ruseski2014}.
 
Third, sport activity was proven to affect social outcomes. For instance, \textcite{caruso2011} showed that sport activity decreases property and juvenile crime among young adults. Moreover, sport activity was found to enfold positive effects on economic outcomes such as wages and earnings \parencites(e.g.)(){lechner2009}{rooth2011}, human capital \parencite{steckenleiter2019}, and quality of work performance \parencite{pronk2004}. Finally, sport activity has been confirmed to lead to higher academic achievements \parencites{fox2010}{pfeifer2010}{felfe2016}{lechner2016}{fricke2018}, to positively affect concentration, memory and classroom behavior \parencite{trudeau2008}, and to improve social relations \parencite{stempel2005}.
 
 The effects of sport activity are, thus, explored in various spheres of human life. The effect of sport activity on human mating, however, is almost unexplored, as discussed next.

\subsection{Sport Activity and Human Mating}

Research on human mating has established in sociology, psychology, economics, and, more recently, computer science, mostly attributable to the range of potential explanatory factors that determine human mating \parencite[][]{eastwick2014} and novel data opportunities due to computer-mediated approaches for human mating (i.e., online dating). In addition to various studies referring to socio-demographic and psychological characteristics to explain human mating \parencite[for a detailed review, see][]{eastwick2014}, a few studies also consider sport activity as potentially relevant factor in explaining human mating.

\textcite{schulte2008} studied the effect of males’ practiced sport discipline on females’ willingness to engage in a relationship, applying an experimental setting. The authors showed that \textit{'[…] team sport athletes were perceived as being more desirable as potential mates than individual sport athletes and non-athletes'} (p. 114). Moreover, the authors argued that \textit{'team sport athletes may have traits associated with good parenting such as cooperation, likeability, and role acceptance'} (p. 114) to explain the positive effect of team sport participation on desirability. However, the authors restrict sport activities to a particular type of sport, namely team vs. individual sport, which, in turn, impedes a valid assessment of the general effect of sport activity on human mating. In a similar vein, \textcite{farthing2005} showed, also applying an experimental setting, that \textit{'[…] females and males preferred heroic sport risk takers as mates, with the preference being stronger for females'} (p. 171) , while interpreting (non-) heroic sport risk as, for example, engaging in (non-) risky sport activities. However, the previously raised concerns apply in the same way to the findings by \textcite{farthing2005}. 

Further research provides insights on potential indirect effects of sport activity on human mating. In particular, previous research indicated that sport activity improves, inter alia, attractiveness \parencite{park2007}, health \parencite{warburton2006}, and income generation \parencite{lechner2009}, all of which have been shown to affect human mating \parencites(e.g.)(){hitsch2010a}{hitsch2010b}{eastwick2014}. However, these studies remain inconclusive with respect to human mating, given the missing integration of relevant context-factors (i.e., further relevant personal/sport characteristics) affecting human mating.

Taken together, sport activity seems relevant for explaining human mating. However, a conclusive, finally valid, assessment on the effect of sport activity on human mating is missing, given that previous research assessed the effect of sport activity on human mating either in the absence of potentially relevant socio-demographic characteristics or by utilizing a narrowed interpretation, respectively representation, of sport activity. These limitations surprise given that information on sport activity are one of the most articulated and visible features on online dating platforms. Furthermore, as discussed previously, sport activity is ascribed relevant effects on various spheres of human life, including physical and mental health as well as social and economic conditions. Following the above mentioned limitations, we focus on the analysis of the effect of sport activity on human mating.

\section{Setup and Data}\label{sec:setup}

In the course of this research, we collaborated with a German online dating platform operator. The operator provided us both with information on the functionality as well as with data from the online dating platform. 

\subsection{Online Dating}
The online dating platform allows a user to virtually meet and communicate with other users. The user has to pay a monthly fixed subscription fee to register and to utilize the online dating platform. The registration at the online dating platform is subdivided into three major sections. First, the user is requested to provide socio-demographic information (e.g., sex, age, education, and income). Second, the user is requested to specify search criteria for potential partners (e.g., sex, age, education, and income). Third, the user is requested to answer a personality test that relates to the users' life style, personality, attitudes and views (79 categories in total). Moreover, the user articulates the language preferences and may include one or more photos on the personal profile page. However, these photos remain fully blurred until the user decides to release the photo for the potential partner.\footnote{In our analysis we restrict the user interactions by excluding the actions involving the release of the blurred photo. We discuss this point in Detail in Section \ref{sec:identification}.} Most importantly, with specific regard to the intended analysis, a user articulates her/his sport preferences and actual sport activities within a total of 27 disciplines, how often she/he actively practices sport, and, finally, which recreational activities dominate in her/his leisure time. A detailed description of the survey questions and the corresponding variables together with descriptive statistics can be found in Online Appendix \ref{App:online}.

Following the registration at the online dating platform, the user can define a query, indicating the preferred sex, age, and geographic location to explore potential partners. The search query returns a shortlist of potential partners, who correspond to the previously defined qualifications. The shortlist includes the potential partners' username, age, a blurred version of the photo, and a matching score, which is computed by the online dating platform operator in order to support users in finding a potentially fitting partner.\footnote{The online dating platform operator does not provide the formula to calculate the matching score. However, it provided us with all data required for its calculation. We elaborate more on this point in Section \ref{sec:identification}.} The user can investigate the potential partner in detail by browsing on the potential partners' profile page, which displays a blurred version of the photo as well as information on the previously described survey. The user can then choose from multiple possible actions. As such, the user can either send a private text message, a 'Smile' icon, or a 'Smile Back' icon (if initially received a 'Smile' icon) to a potential partner. Additionally, a user may leave a 'like' or a text note on a potential partners' profile page. Moreover, the user can initiate a friendship with a potential partner by initiating a profile release or accepting an initial profile release by a potential partner. Furthermore, a user may request an 'Applet' (game with questions) to a potential partner, which works out similarities/differences between the user and the potential partner. Finally, a user may prevent unwanted users from contacting in any form.

\subsection{Data}
The acquired data consists of two samples. The first, \textit{user sample}, contains personal information about the registered users on the platform. The second, \textit{interaction sample}, contains information about the users' interactions on the platform.

The user sample includes 18'036 newly registered users who joined the platform between January 1st, 2016 and April 30th, 2016.\footnote{Other empirical studies using online dating data focused on observation periods of similar length \parencites[see][and]{hitsch2010a}{hitsch2010b}.} For each registered user, we observe the full information filled upon the registration, which comprises 667 variables in total. For our intended analysis with regard to the sport activity, we exclude the users with daily sport frequency, as these comprise only around 3\% of all users, which would prevent a meaningful analysis for this group. Furthermore, we restrict ourselves to the sample of users, whose residency is located in Germany, as only for these users we observe full location information, including the ZIP codes. This restriction affects only about 2\% of the observations as the platform provider operates on the German market. Lastly, we exclude users with incomplete information (around 1\% of the sample) and those with implausible and inconsistent values (less than 1\% of the sample).\footnote{This includes, for example, users with more than one single value for a mutually exclusive answer selection, among others.} This leaves us with an available sample consisting of 16'864 users for our analysis. A descriptive summary of selected variables for the user sample is presented in Appendix \ref{App:B}.

The interaction sample includes 1'415'645 user actions among the population of newly registered users over the same time period. For each action, we observe the IDs of both users involved in the action, as well as the precise time stamp and the type of action. Each interaction between users must begin with a visit action (invisible to a user), upon which further types of actions are possible, such as a message, like or smile (visible to a user). We refer to the user who initiates an interaction as a \textit{sender} of an action, and the user who gets involved in an interaction as a \textit{recipient} of an action. For the purposes of our analysis, we filter the interactions such that we consider only \textit{one-way} interactions initiated by a visit action, with either no further action at all, or immediately followed by any visible action from the sender, without considering any visible recipient's response to the initial action from the sender. Thus, we select only unique interactions in the sense that the sender was visibly or invisibly active, while the recipient stayed visibly passive. Thereby, we restrict the interactions between the users until the point of a possible reciprocal interaction taking place.\footnote{For a more detailed definition of valid user interactions with practical examples, see Appendix \ref{App:A}.} This selection of the sample will be later important for the validity of our identification strategy (see Section \ref{sec:identification} for details). We further shape our sample such that each observation represents a valid interaction accompanied by indicators of visible sender actions that have taken place within the particular interaction as well as the sender and recipient user IDs. This leaves us with an available sample consisting of 178'372 valid unique interactions for our analysis. 

Lastly, to construct our final estimation sample, we merge the interaction sample with the user sample. As a result, each observation in our estimation sample represents a valid interaction between two users and consists of sender and recipient user IDs together with sender's actions from the interaction sample, and both the sender's as well as recipient's characteristics obtained from the user sample. Furthermore, as the data contains only heterosexual users based on a binary measure for gender, i.e. we never observe a sender and recipient of the same sex in our sample, we split the sample based on gender for a clearer interpretation of the results. Hence, we refer to the sample with only female recipients as the \textit{female sample}, as here the females are in the role of an approached user upon receiving a visit action, and possibly further actions, by a male sender of an action. Analogously, we refer to the sample with only male recipients as the \textit{male sample}, as in this case the males are in the role of an approached user upon receiving a visit action, and possibly further actions, by a female sender of an action.

Thus, we are left with 108'456 observations for the female sample and with 69'916 observations for the male sample. The corresponding descriptive statistics of selected variables for the two samples are listed in Appendix \ref{App:B}.

\subsubsection{Sport Activity}
In order to investigate the effect of sport in online dating, we leverage the rich information set regarding the sport activities on the user profile. In particular, each profile includes a detailed statement of the user's sport frequency. This information stems from the initial questionnaire filled by the user upon registration. First, the user is asked about the sport types done actively, namely: \textit{'What sports do you do actively?'}, with multiple options (mutually inclusive) such as basketball, fitness, hiking, soccer, tennis, etc., or specifying the option \textit{'none'}. Second, only if the user has not specified the option \textit{'none'}, a further question
regarding the particular sport frequency is asked: \textit{'How often do you practice sport?'}. The possible values (mutually exclusive) include the following answers: \textit{'every day', 'several times a week', 'several times a month',} or \textit{'less common'}. Thus, we not only observe the user's binary indication of practicing sport or not, i.e. the extensive margin, but also the particular sport frequency, i.e. the intensive margin. This provides us with a much finer measure of the actual sport activity. Accordingly, we define the sport activity measure to be multi-valued with sport frequencies of \textit{weekly, monthly, rarely} and \textit{never}. We omit the daily frequency for lack of data within this category, as previously mentioned. Furthermore, we leave the sport types out of consideration too, as these include many different and not mutually exclusive values, which prevents a clear separation of the categories.

Table \ref{tab:treatment} shows the descriptive statistics for the sport frequency shares in the samples of males and females, respectively as well as the corresponding shares from the innovation sample of the German socioeconomic panel \parencite[SOEP-IS;][]{soep2015} for a comparison with a representative population sample.\footnote{Similar values for the sport frequency statistics for Germany are documented also in the Eurobarometer Survey \parencite{eurobarometer2014}, as pointed out by \textcite{steckenleiter2019}.} First, we see that the sport frequency is unevenly distributed in both samples. Second, we can also observe that the shares are very similar in both samples. Nonetheless, the \textit{never} category is more represented in the female sample, while the \textit{weekly} category is more represented in the male sample. Additionally, we also observe that the subjective sport frequency of the users from the online dating platform is in general much higher than the one of the representative individuals from Germany.\footnote{Note, that this might be both due to truly higher sport frequency of the registered users as well as due to an overestimation of own actual sport frequency of the users, or the combination of both. Also note, that our sample consists only of singles, which is in contrast to the representative population sample.} Third, with respect to the number of observations in the corresponding samples, we immediately see that even though we have a balanced user sample in terms of gender,\footnote{The user sample consists of 48\% of females and 52\% of males. For more descriptive statistics with regard to the user sample, see Appendix \ref{App:B}.} females get visited more often than males do.

\begin{table}[H]
\centering
\caption{Shares of Sport Frequency for Male and Female Sample}\label{tab:treatment} 
\begin{tabular}{rrrrrr}
  \toprule
 & \cellcolor{white} \textit{Never} & \cellcolor{white} \textit{Rarely} & \cellcolor{white} \textit{Monthly} & \cellcolor{white} \textit{Weekly} & Observations \\ 
  \midrule
   Males & \cellcolor{malegreen!15}0.07 & \cellcolor{malegreen!25}0.08 & \cellcolor{malegreen!45}0.29 & \cellcolor{malegreen!65}0.56 & 69'916 \\ 
   Females & \cellcolor{femalegreen!25}0.12 & \cellcolor{femalegreen!15}0.09 & \cellcolor{femalegreen!45}0.29 & \cellcolor{femalegreen!65}0.49 & 108'456 \\
   \hline
   SOEP-IS & 0.33 & 0.23 & 0.10 & 0.34 & 25'544 \\ 
   \bottomrule
   \multicolumn{6}{l}{\footnotesize \textit{Note:} Color intensity represents the corresponding share sizes for males and females.}\\
\end{tabular}
\end{table}

\vspace{-0.2cm}

Finally, given our definition, the impact of sport activity can be illustrated as follows. The user, here the sender, visits a profile of another user, here the recipient, and gets exposed to an information revealed on the profile. Among other indicators, the sender observes the recipient's indication of the sport frequency, i.e. the variable of interest. Based on the available information, the sender then decides to perform or not to perform a further action.

\subsubsection{The Interaction between Users}
We are interested in the one-way actions of a sender upon visiting a recipient's profile on the website. Even though there are multiple actions a sender can initiate, we focus explicitly on the action of sending a text message for several reasons. First, a text message is the most evident action of showing a serious interest, as in order to compose a text message, the sender has to exhibit a substantial effort, in comparison to other available options, such as simply sending a smile or like. Second, unlike the other generic options, by sending a text message, the sender directly approaches the recipient in an individualized manner. Third, an outcome measure of sending a text message or an email has been previously used in the online dating literature under the assumption that users send a message if and only if the potential utility of the match exceeds some minimum threshold value \parencites(compare e.g.)(){hitsch2010a}[or][]{bruch2016}. Hence, we define our action of interest as a binary measure of sending (1) or not sending (0) a text message upon a profile visit. Given the binary scale, the natural interpretation as contact chances in terms of message probabilities arises.

\begin{figure}[H]
    \centering
    \caption{Average Contact Chances according to Sport Activity for Males and Females}
    \label{fig:outcome}
    \includegraphics[width=0.95\textwidth]{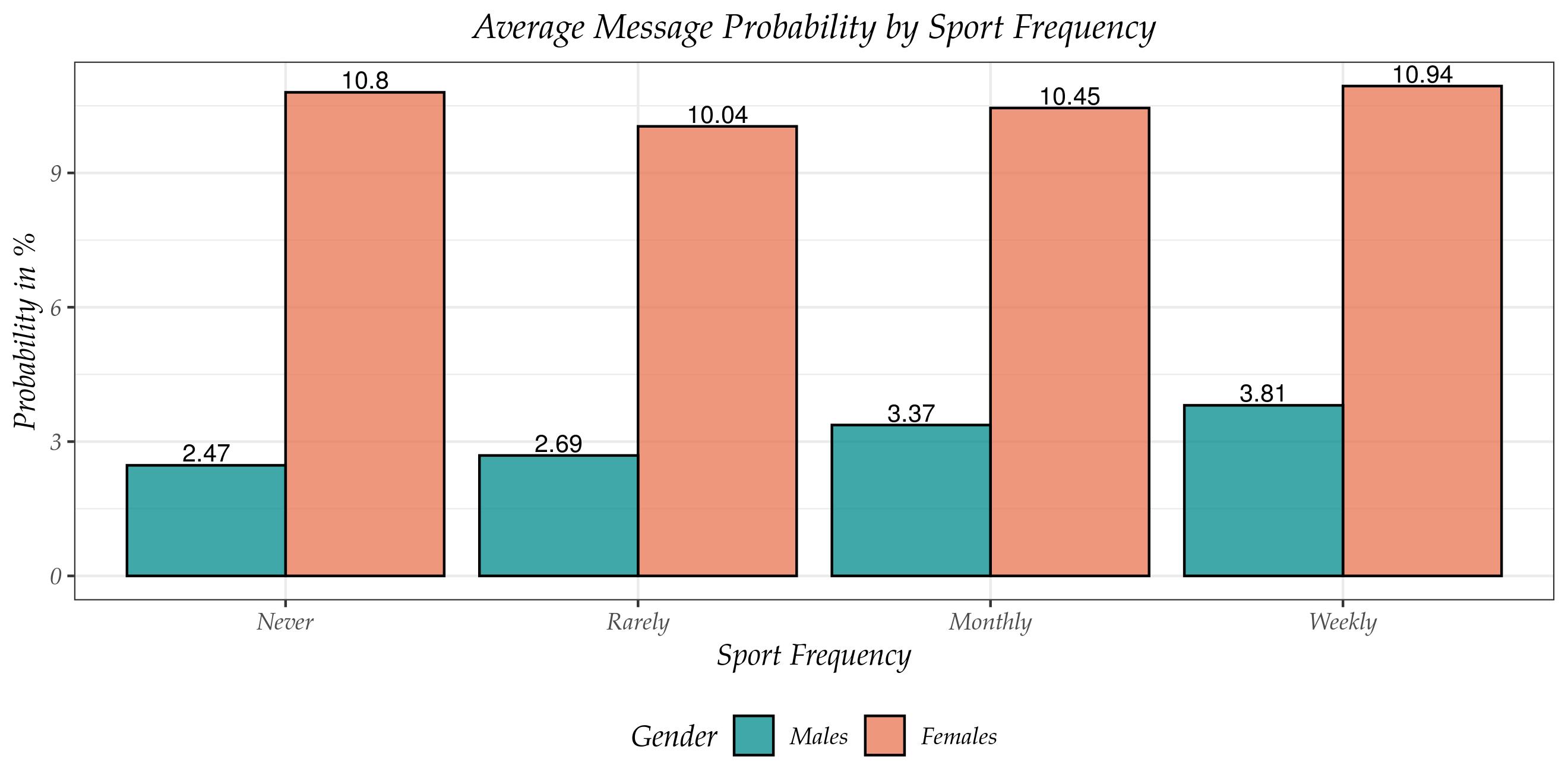}
\end{figure}

\vspace{-0.25cm}

Figure \ref{fig:outcome} shows the average message probability in percentages for males and females according to the sport frequency. First, we see that the levels of females are substantially higher than those of males, i.e. women have unconditionally a higher probability to get messaged than men do. This is in line with previous evidence from studies based on online dating data \parencite{bruch2016}. Second, we observe a slightly increasing message probability with higher sport frequency for males, while for females no clear pattern can be identified.

\subsubsection{Information about Users}
In our sample, we have access to complete information filled by the user upon registration. Hence, we not only observe the condensed information displayed on the main user profile page, but also the expanded information stored in the background of the user profile. Thus, we effectively observe the very same information that a real user observes upon a profile visit of a potential partner, and even beyond. The full information observable to us includes the following components. First, we observe the user's demographic information such as gender, age, height, etc., the socio-demographic information such as education and income level, type of occupation, etc., as well as personal information such as place of residence, smoking habits, or even (self-judged) appearance. Second, in addition to the user specific information, we observe the user's preferences for a potential partner in terms of the search criteria related to the above mentioned socio-demographic information as well. Third, we furthermore observe the user's information stemming from the detailed personality test, which reflects on the user's life style, personality, attitudes and preferences. This includes an extensive information on topics like religion, political views, music and travel preferences, or even partner requirements. The aforementioned user information comprises of an exhaustive list of 663 variables in total. However, given the structure of our data, we include the user information both for the recipient as well as for the sender, resulting in effectively more than thousand variables. Apart from the information coming directly from the platform, we additionally generate a variable measuring the distance between the recipient and the sender, based on the available ZIP codes.\footnote{The average distance between the recipient and the sender in our sample is 67.32 km. A detailed plot of the distribution of the distance between the users ca be found in Appendix \ref{App:B}.}

We consider all the aforementioned variables as controls in the sense of potential confounders, i.e. as variables jointly influencing both the recipient's sport activity as well as the recipient's potential outcome of receiving or not receiving a text message, and thus, \textit{de facto} the sender's action to contact or not to contact the recipient. Conditioning on such a large-dimensional covariate space is a challenging estimation task. However, we refrain ourselves from an arbitrary selection of the confounding variables in order to reduce the dimension of the estimation problem. Rather, we apply a novel causal machine learning estimator, which can effectively deal with such large-dimensional setting, performing implicit variable selection in a flexible and data-driven way.
The only variable deselection we perform manually is related to endogenous variables.\footnote{We elaborate on this issue more closely when discussing the identifying assumptions in Section \ref{sec:identification}.} Thus, we remove all variables that could be potentially influenced by the sport frequency. These include mainly variables indicating the specific sport type, but also variables describing sport-related choices such as holiday and leisure time preferences, as well as variables regarding the body type and clothing style. In total, we dismiss 38 endogenous variables. Lastly, we leave out 2 variables without any variation. As a result, we are left with 1247 covariates in total (1229 ordered, including dummies and 18 unordered), reflecting the recipient and sender characteristics.

Apart from the confounding role, the covariates are useful for analysing the effect heterogeneity, too. For this purpose, we pre-specify a small subset of heterogeneity variables, consisting of age, income and education level on both recipient as well as sender side, together with the corresponding distance between the recipient and the sender. We focus on these heterogeneity variables for two main reasons. First, these socio-demographic information are widely recognized in the literature as being the main determinants of the partner choice (for a review of the importance of selected socio-demographic characteristics see \cite{hitsch2010a} and \cite{eastwick2014}). Second, these are also the main variables that are most visible to the user on the profile summary and thus can potentially impact the shape of the effect. Additionally, we analyze the heterogeneous effects also along the sport frequency for the recipient as well as for the sender. Complementary to the pre-specified subset of heterogeneity variables, the remaining variables might serve for a supplementary descriptive analysis of the effects. 

\section{Empirical Approach}\label{sec:empirics}
To analyze the effect of sport activity on human mating, we leverage the recent advances in the causal machine learning literature. Below, we outline the parameters of interest together with the identification and estimation thereof.

\subsection{Parameters of Interest}
In order to define the parameters of interest, we rely on the Rubin's \parencite*{rubin1974} potential outcome framework. We denote the treatment variable of a user $i$ by $D_i$, which in our case can take on four different integer values, i.e. $D_i \in \{0,1,2,3\}$, corresponding to sport frequencies of \textit{never, rarely, monthly,} and \textit{weekly}, respectively. According to the treatment status, $d$, we define the potential outcomes for the user $i$ by $Y_i^d$, which in this case is the action of receiving or not receiving a text message. However, we only observe the potential outcome under the treatment which the user $i$ is associated with \parencites[see][for a discussion of the fundamental problem of causal inference]{holland1986}. Thus, the realized outcome can be defined through the observational rule as follows: $Y_i=\sum^3_{d=0}\mathbb{I}(D_i=d)\cdot Y_i^d$, which implies that we observe the action of receiving the text message only under a particular sport frequency of the recipient. Further, we denote the observed vector of covariates by $X_i$, which contains the recipient and sender characteristics, together with a subset of pre-specified heterogeneity variables $Z_i$, such that $Z_i \subset X_i$.

To analyze the effect of sport frequency on the message probability, we are interested in the following causal parameters. First, the \textit{Average Treatment Effect (ATE)} of treatment $D_i=m$ compared to treatment $D_i=l$ is defined as $$ATE=\theta=\mathbb{E}[Y_i^m - Y_i^l]$$ and constitutes the classical parameter of interest in microeconometrics, which provides us with an aggregated effect measure \parencite[compare e.g.][]{imbens2009}. Second, the \textit{Group Average Treatment Effect (GATE)} is characterized as $$GATE=\theta(z)=\mathbb{E}[Y_i^m - Y_i^l \mid Z_i=z]$$ and measures the differential effects along the heterogeneity variables $Z_i$. Thus, it provides us with a disaggregated effect measure according to the specific variables of interest, as in our case is the age, income and education level, distance as well as the sport frequency itself. In the latter case, the GATE corresponds to the \textit{Average Treatment Effect on the Treated (ATET)}. Third, the \textit{Individualized Average Treatment Effect (IATE)} is denoted as $$IATE=\theta(x)=\mathbb{E}[Y_i^m - Y_i^l \mid X_i=x]$$ and describes the heterogeneous effects based on the full set of observed covariates $X_i$. As such, the IATEs present the disaggregated effects on the finest level of granularity and thus provide us with user-type specific effects.

Notice, that both the treatment variable, i.e. the sport frequency, as well as the outcome variable, i.e. receiving a text message, are measured on the recipient side, and hence, also the above defined causal effects refer to the recipient.

\subsection{Identification Strategy}\label{sec:identification}
Given our observational study design, it is not possible to only compare the unconditional message probabilities for different sport frequencies, as displayed in Figure \ref{fig:outcome}, to infer the causal effects, since the user decision regarding the sport activity is not random. The level of sport frequency might be influenced by other variables representing socio-demographic information, which might also influence the potential outcome of receiving or not receiving a text message. For example, recipients with a higher level of education might have a higher probability of doing sport on a weekly basis, as well as a higher probability of getting messaged. This phenomenon is known as selection bias \parencite{imbens2009}. In order to disentangle the causal effect from the selection effect, we need to eliminate such confounding via credible identification strategy.

For the identification of the aforementioned parameters of interest in a multiple treatment case, we rely on the so-called \textit{selection-on-observables} strategy \parencites(see)(){imbens2000}[or][]{lechner2001}. Such identification approach assumes that all confounding variables jointly influencing both the treatment as well as the potential outcomes are observed, and thus, can be conditioned on. Given our rich data on user characteristics and the unique research design, we argue to capture all possible confounding effects for two main reasons. First, for both the recipient and the sender, we observe socio-demographic (e.g., age, education, income) and personal (e.g., family status, smoking habits, place of residence) characteristics, together with the preferences for a potential partner as well as the answers given in a detailed personality test. Thereby we have access to even richer personal information than the actual users when browsing the profiles, and as such, we are able to control for confounding effects stemming from the user's characteristics. Second, given our research design, focusing only on the very first one-way interactions between the recipient and the sender, we effectively eliminate any possible unobserved effects coming from the reciprocal interaction between the users such as sympathy or kindness. By doing so, we explicitly focus only on situations, in which the recipient's profile gets visited by a sender, upon which the recipient does receive or does not receive the very first text message from the sender, without any visible encouragement to do so from the recipient her/him-self. In such a situation, the sender decides solely based on the information visible on the recipient's profile to send or not to send the message. Within our research design, we observe exactly the same information, and even beyond, as the actual sender when facing the decision of sending the first text message. For this reason, we are also able to control for confounding effects stemming from the user's interaction. 

Taken together, combining the highly-detailed user information, which exceeds the information directly observable by the actual users, with the unique research design, which eliminates any possible unobservable information, we are confident to capture all confounding effects. In particular, our selection-on-observables strategy relies on the following set of identification assumptions.

First, the so-called \textit{conditional independence} assumption (CIA), states that the potential outcomes and the treatment are independent once conditioned on the covariates. This hinges on the availability of all covariates that jointly influence the potential outcome and the treatment. As we argue, we observe sufficiently rich information on both the recipient as well as the sender side to ensure the plausibility of the CIA. In addition, our research design eliminates any further influence from a possible reciprocal interaction between the users. Thus, we are confident about the validity of the CIA in this particular case. There are only two potential sources of vulnerability of the CIA in this case. First, it could be caused by the availability of the blurred photo of the user. Even though the photo remains blurred, as we do not allow interactions between the users which would include the action to release the photo, we cannot rule out that information such as the shape of the face or the hair and skin colour could be, nonetheless, inferred. However, despite the information inferred from the blurred photo might possibly affect the outcome, i.e. the message probability of the recipient, we argue that this information should not have an effect on the treatment itself, i.e. the recipient's sport frequency. Thus, it arguably does not qualify as a potential confounder. Nevertheless, limitations in the availability of profile pictures, respectively opportunities to represent the information in profile pictures, are common in the literature on online dating \parencite{fiore2008}. Second, it could be caused by the availability of the matching score. However, despite the fact, that we do not observe the score directly, we know that we observe, and indeed condition on, all information which serves for its calculation. Moreover, even though we do not know the exact formula, by using a very flexible estimation approach, we are able to reproduce any arbitrary functional form of the matching score. Nonetheless, if the matching score would consist of the user's sport frequency, the treatment would be indirectly observed as a part of the shortlist of potential partners even before actually visiting the user profile. However, this would not violate the CIA as such, it could rather potentially reduce the size of our effect estimates. For this reason, we conduct a placebo test to provide evidence that this is indeed not the case. We discuss the placebo test in more detail in Section \ref{sec:placebo}.

Second, the \textit{common support} assumption, ensures that for each value in the support of the covariates, there is a possibility to observe all treatments. This means that we find users with the same age, education, income, etc., for all sport frequency levels. Thus, we are able to check the validity of the common support assumption in the data directly, but do not find any violations thereof \parencite[see][for a discussion of common support issues]{lechner2019}.

Third, the \textit{stable unit treatment value} assumption \parencite[see e.g.][]{rubin1991}, implies that for each user we observe only one of the potential outcomes based on the treatment status. It further implies that there is no interference among users, hence ruling out any general equilibrium or spillover effects. This means that the sport frequency of one particular user does not affect the message probability of other users. We argue that the SUTVA is plausible in this case, as we analyze only a short time period after the user registration such that general equilibrium or learning effects would not yet emerge.

Fourth, the \textit{exogeneity of confounders} assumptions, indicates that the values of the covariates are not influenced by the treatment. In other words, the user characteristics should not be impacted by the sport frequency. For this reason, we discard all potentially endogenous variables such as indicators of particular sport type, sporty clothing style, preferences for sport holidays or sport club memberships. Therefore, we are confident that the exogeneity assumption holds.

Under the aforementioned assumptions, it can be shown that the above parameters of interest are identified. For technical details, see \textcite{lechner2018}.

\subsection{Estimation Method}
In our analysis, we face two major challenges with regard to the estimation of the causal effects of interest. First, we need to deal with a very large conditioning set with an unknown functional form of the covariates. Second, we want to investigate potential effect heterogeneity. In order to overcome these challenges, we take advantage of the newly developing causal machine learning literature \parencites(see)(){athey2018}[][]{athey2019}[or][for overviews]{knaus2020}. It combines
the flexibility and prediction power of machine learning \parencite{hastie2009} with the causal inference from econometrics \parencite{imbens2009}. One of the most popular machine learning methods are the so-called regression trees \parencite{breiman1984} and random forests \parencite{breiman2001}. The trees and forests are highly flexible, local nonparametric prediction methods, which can effectively deal with large-dimensional settings \parencite{biau2016}. Adapting these prediction algorithms towards causal inference has lead to developments of Causal Trees \parencite{athey2016} and Causal Forests \parencite{wager2018}, respectively. These methods inherit the advantages of the prediction versions, while flexibly estimating the causal effects with systematically uncovering their heterogeneity. Furthermore, Lechner \parencite*{lechner2018} extends the Causal Forest for the multiple treatment case, and additionally improves the splitting rule to account for selection bias and for the mean correlated error. The resulting Modified Causal Forest also allows for estimation as well as inference for the parameters of interest at all aggregation levels in one estimation step. Since our application involves multiple treatments with potential confounding, while analyzing various heterogeneity levels of the causal effects, we opt for the latter approach.

In our analysis, we rely on estimating the so-called 'honest' forest, which has been shown to lower the bias of the causal effect estimates and to enable valid statistical inference \parencites[][and]{wager2018}{lechner2018}. As such, we randomly split the estimation sample in two equally sized parts and use one sample, i.e. the \textit{training} sample, to build the Modified Causal Forest and the other sample, i.e. the \textit{honest} sample, to estimate the causal effects.\footnote{Lechner \parencite*{lechner2018} shows in a simulation study that the efficiency loss of the 'honest' forest due to sample-splitting is minimal in comparison to the case of 'honest' trees as in Wager and Athey \parencite*{wager2018}.} Then, the estimation procedure of the Modified Causal Forest can be described as follows. First, the estimator draws a random subsample $s$ of the training sample and subsequently estimates a single causal tree. As such, the subsample gets recursively splitted into smaller subsets, the so-called 'leaves' of the tree $L(x)$ . The partitioning follows a splitting rule which removes selection bias and reveals effect heterogeneity. As a result, the observations are homogeneous with regard to the covariate values \textit{within} the leaf, while being heterogeneous \textit{across} the leaves. Then, the treatment effect is estimated within each terminal leaf by simply subtracting the mean outcomes of the respective treatment levels $D_i=m$ and $D_i=l$ from the honest sample as

$$\hspace{-0.25cm}\hat{\theta}_s(x)=\frac{1}{\{ i: D_i=m, X_i \in L(x) \}}\hspace{1cm}\sum_{\mathclap{\{ i: D_i=m, X_i \in L(x) \}}}Y_i \hspace{0.5cm}- \hspace{0.25cm}
\frac{1}{\{ i: D_i=l, X_i \in L(x) \}}\hspace{1cm}\sum_{\mathclap{\{ i: D_i=l, X_i \in L(x) \}}} Y_i.$$

Second, as a single tree might be quite unstable due to its path-dependent nature, the forest estimates many such trees by drawing $S$ random subsamples in total. The Causal Forest estimate is then given by the ensemble of many causal trees as

$$\widehat{IATE}=\hat{\theta}(x)=\frac{1}{S}\sum_{s=1}^S \hat{\theta}_s(x).$$

The additional averaging of the trees helps to reduce the variance and to smooth the edges of the leaves \parencite{buhlmann2002}. Conceptionally, the Causal Forest can be thought of as a nearest neighbor matching estimator with an adaptive neighbor choice and can be thus described using a weighted representation, too \parencites{wager2018}{wager2019}. 

Third, the Modified Causal Forest estimates the GATEs by averaging the IATEs in the corresponding subsets defined by the heterogeneity variables $Z_i$ and the ATE by averaging the IATEs in the whole sample as follows

$$\widehat{GATE}=\hat{\theta}(z)=\frac{1}{\{i: Z_i=z \}} \sum_{\{i: Z_i=z \}} \hat{\theta}(x)$$

and

$$\widehat{ATE}=\hat{\theta}=\frac{1}{N}\sum_{i=1}^N \hat{\theta}(x).$$

Thus, it provides a computationally attractive option to estimate the effects of interest on all desired levels of heterogeneity without the need for re-estimating the whole forest for each single aggregation level.\footnote{In our setting, we additionally apply treatment sampling probability weights for the ATE and GATEs aggregation of the IATEs to account for the unbalanced treatment shares.}

Fourth, the Modified Causal Forest then explicitly uses the weighted representation of the estimated effects for inference. The weight-based inference can be then conveniently applied to all aggregation levels as well.\footnote{\textcite{wager2019} further suggest usage of the forest weights for solving many different econometric estimation problems.} For an in-depth discussion of the Modified Causal Forest, see \textcite{lechner2018} as well as \textcite{cockx2020} and \textcite{hodler2020} for empirical applications.

Estimating the effects of sport frequency on the message probability by applying causal machine learning allows us to improve on previous empirical studies in an online dating setting in several dimensions. First of all, we do not have to specify the exact functional relationship between outcome, treatment and covariates, as in the case of using parametric approaches such as the logistic regression \parencites(see e.g.)(){hitsch2010a}{hitsch2010b}[or][]{bruch2016}. This is particularly important when dealing with a large-dimensional covariate space, including the characteristics of both the recipient and the sender, as the functional form of the interactions thereof is not  \textit{a priori} clear. Furthermore, using causal machine learning also advances the semiparametric approaches used in online dating studies \parencite[see e.g.][for a matching estimation]{lee2016}, thanks to more flexible adaptive estimation and its implicit variable selection properties. Lastly, causal machine learning allows us to go beyond the average effects and systematically investigate the effect heterogeneity on various aggregation levels, without the need to specify interactions or to build subsets of data in an \textit{ad-hoc} fashion.

\section{Results}\label{sec:results}
Below, we present the results for the average and heterogeneous effects of sport activity on contact chances, based on the Modified Causal Forest estimation.

\subsection{Average Effects}
The results for the average effects of the sport activity on the contact chances are summarized in Table \ref{tab:ate}. The diagonal presents the potential outcomes, while the corresponding effects are depicted in the lower triangle.

In case of the male sample, for increasing sport frequency, the results show a clear and increasing pattern of the potential outcomes, i.e. of the potential message probability. While the potential message probability for users who never practice sport is on average only $2.50\%$, for users doing sport on a weekly basis, the chances to get messaged increase by more than $50\%$ and amount to $3.82\%$. Comparing the respective potential outcomes across the sport frequency levels yields the corresponding causal effects measured in percentage points. Accordingly, all effects for all sport frequency comparisons are positive. The most sizeable and the most precise effects are estimated for the most distinct sport frequencies, as one would intuitively expect. Thus, the average effect of a weekly sport activity versus no sport activity at all, is equal to an 1.32 percentage points increase. Similarly, the average effect of a weekly in comparison to only rare sport activity amounts to an 1.20 percentage point increase. Moreover, these effects are both substantively as well as statistically relevant. As such, a male user increasing his sport activity from no sport or only rare sport activity to doing sport on a weekly basis significantly increases the probability of getting messaged by $52.80\%$ and $45.80\%$, respectively. In practice, this implies receiving 13, respectively, 12 extra messages out of 1000 profile visits. Hence, the contact chances of a male user can be substantially increased solely by becoming more sporty. The remaining effects comparing less distinct sport frequencies lack the statistical relevance, which stems mainly from the substantially lower number of observations for these categories (see Table \ref{tab:treatment}).

Regarding the female sample, the results do not suggest increasing contact chances with increasing frequency of sport activity, as in the case of the male sample. The potential outcomes thus do not indicate any clear pattern as the message probability firstly drops, when switching from no sport to rare sport activity, and then increases steadily throughout the monthly and weekly sport frequencies, reaching comparable levels with the category of never doing sport. Accordingly, the estimated average effects do not show any explicit structure and lack statistical relevance. The only exception is the precise estimate of the effect of the weekly vs. rare sport activity, with a sizeable increase of an 1.61 percentage points, yet this represents only a minor relative increase of $17.18\%$ in comparison to the effects seen in the male sample. Taken together, based on the overall results, no substantial conclusions can be drawn.

\begin{table}[H]
\centering
\caption{Average Effects of Sport Activity on the Contact Chances for Males and Females}\label{tab:ate}
\resizebox{1.0\textwidth}{!}{%
\begin{tabular}{lp{1.1cm}p{1.1cm}p{1.1cm}p{1.1cm}p{1.1cm}p{1.1cm}p{1.1cm}p{1.1cm}p{1.1cm}}
  \toprule
 & \multicolumn{4}{c}{Males} & & \multicolumn{4}{c}{Females} \\
  \midrule
 &  \cellcolor{white}\textit{Never} &  \cellcolor{white}\textit{Rarely} &  \cellcolor{white}\textit{Monthly} &  \cellcolor{white}\textit{Weekly} &  & \cellcolor{white}\textit{Never} &  \cellcolor{white}\textit{Rarely} &  \cellcolor{white}\textit{Monthly} & \cellcolor{white}\textit{Weekly} \\ 
  \midrule
\cellcolor{white}\textit{Never} & \cellcolor{malegreen!15}2.50 &  &  &  &  & \cellcolor{femalegreen!75}10.67 &  &  &  \\ 
   & \cellcolor{malegreen!15}\scalebox{0.75}{(0.46)} &  &  &  &  & \cellcolor{femalegreen!75}\scalebox{0.75}{(0.60)} &  &  &  \\ 
\cellcolor{white}\textit{Rarely} & 0.12 & \cellcolor{malegreen!35}2.62 &  &  &  & -1.30 & \cellcolor{femalegreen!35}9.37 &  &  \\ 
   & \scalebox{0.75}{(0.63)} & \cellcolor{malegreen!35}\scalebox{0.75}{(0.43)} &  &  &  & \scalebox{0.75}{(0.88)} & \cellcolor{femalegreen!35}\scalebox{0.75}{(0.63)} &  &  \\ 
\cellcolor{white}\textit{Monthly} & 0.86 & 0.74 & \cellcolor{malegreen!55}3.36 &  &  & -0.29 & 1.01 & \cellcolor{femalegreen!55}10.38 &  \\ 
   & \scalebox{0.75}{(0.52)} & \scalebox{0.75}{(0.50)} & \cellcolor{malegreen!55}\scalebox{0.75}{(0.26)} &  &  & \scalebox{0.75}{(0.71)} & \scalebox{0.75}{(0.72)} & \cellcolor{femalegreen!55}\scalebox{0.75}{(0.40)} &  \\ 
\cellcolor{white}\textit{Weekly} & \cellcolor{white}\textbf{1.32}$^{***}$ & \cellcolor{white}\textbf{1.20}$^{**}$ & 0.46 & \cellcolor{malegreen!75}3.82 &  & 0.31 & \cellcolor{white}\textbf{1.61}$^{**}$ & 0.60 & \cellcolor{femalegreen!95}10.98 \\ 
  & \cellcolor{white}\scalebox{0.75}{(0.50)} & \cellcolor{white}\scalebox{0.75}{(0.47)} & \scalebox{0.75}{(0.32)} & \cellcolor{malegreen!75}\scalebox{0.75}{(0.19)} &  & \scalebox{0.75}{(0.67)} & \cellcolor{white}\scalebox{0.75}{(0.68)} & \scalebox{0.75}{(0.45)} & \cellcolor{femalegreen!95}\scalebox{0.75}{(0.30)} \\ 
   \bottomrule
   \multicolumn{10}{l}{\footnotesize \textit{Note:} Effects in $\%$ points. Potential outcomes on the diagonal. Standard errors in parentheses. Significance levels}\\
   \multicolumn{10}{l}{\footnotesize refer to: $^{***} < 0.01$, $^{**} < 0.05$, $^{*} < 0.1$. Color intensity represents the corresponding level sizes.}
\end{tabular}%
}
\end{table}

\vspace{-0.2cm}

In general, based on the results of the average effects, we find sizeable and significant positive effects of a more frequent sport activity, when analyzing the male users, while we find only weak evidence for such effects for the case of female users. It means that for men a higher sport frequency substantially increases the probability of getting messaged by a woman, on average. However, higher sport frequency for women does not seem to consistently lead on average to considerably higher chances of getting messaged by a man.

\subsection{Heterogeneous Effects}
While the average effects provide a general measure for the causal effects of sport activity, a more detailed description of the effect heterogeneity beyond gender, remains unknown. Therefore, we study the heterogeneous effects in respect to the pre-defined set of variables of interest, i.e. the group average treatment effects (GATEs), to uncover possibly differential effects of the sport activity on the contact chances. For the sake of clarity, we focus on the effects comparing the most distinct cases, namely the weekly sport frequency with no sport activity. In this regard, we analyze the effect heterogeneity along age, education and income of the users, on both the recipient as well as the sender side, together with the mutual user distance, based on the following considerations. First, these variables have been previously identified as the main determinants of the partner choice \parencites{hitsch2010a}{eastwick2014} and second, these are also the variables which appear on the main profile summary. Thus, we expect these variables to have a higher potential to influence the shape of the effect of the sport activity. Lastly, we investigate the effect heterogeneity based on the particular recipient's as well as sender's sport frequency, which is a natural choice as it corresponds to the effect on the treated, a classical microeconometric parameter of interest \parencite[compare e.g.][]{abadie2018}.
Essentially, the heterogeneity analysis enables us to investigate if the benefits of the regular sport activity in terms of higher contact chances vary among specific groups of users. Thus, we shed light on the open questions such as if potentially the users with higher age, or with lower education and income level enjoy higher benefits of weekly sport activity than those with lower age, or with higher education and income level, or vice versa.

In order to test for the presence of heterogeneity along the variables of interest, we conduct the Wald test of equality of the estimated GATEs. Additionally, we conduct $t$-tests for differences of the estimated GATEs from the average effect. Rejection of both tests thus gives support for the existence of heterogeneity with respect to the particular variable.\footnote{Detailed results of the Wald test for equality of the GATEs as well as the tests for differences from the ATE are listed in Appendix \ref{App:D}.}

\begin{figure}[H]
    \centering
    \caption{Heterogeneous Effects of Sport Activity based on Income for Males}
    \label{fig:male_gates}
    \includegraphics[width=0.95\textwidth]{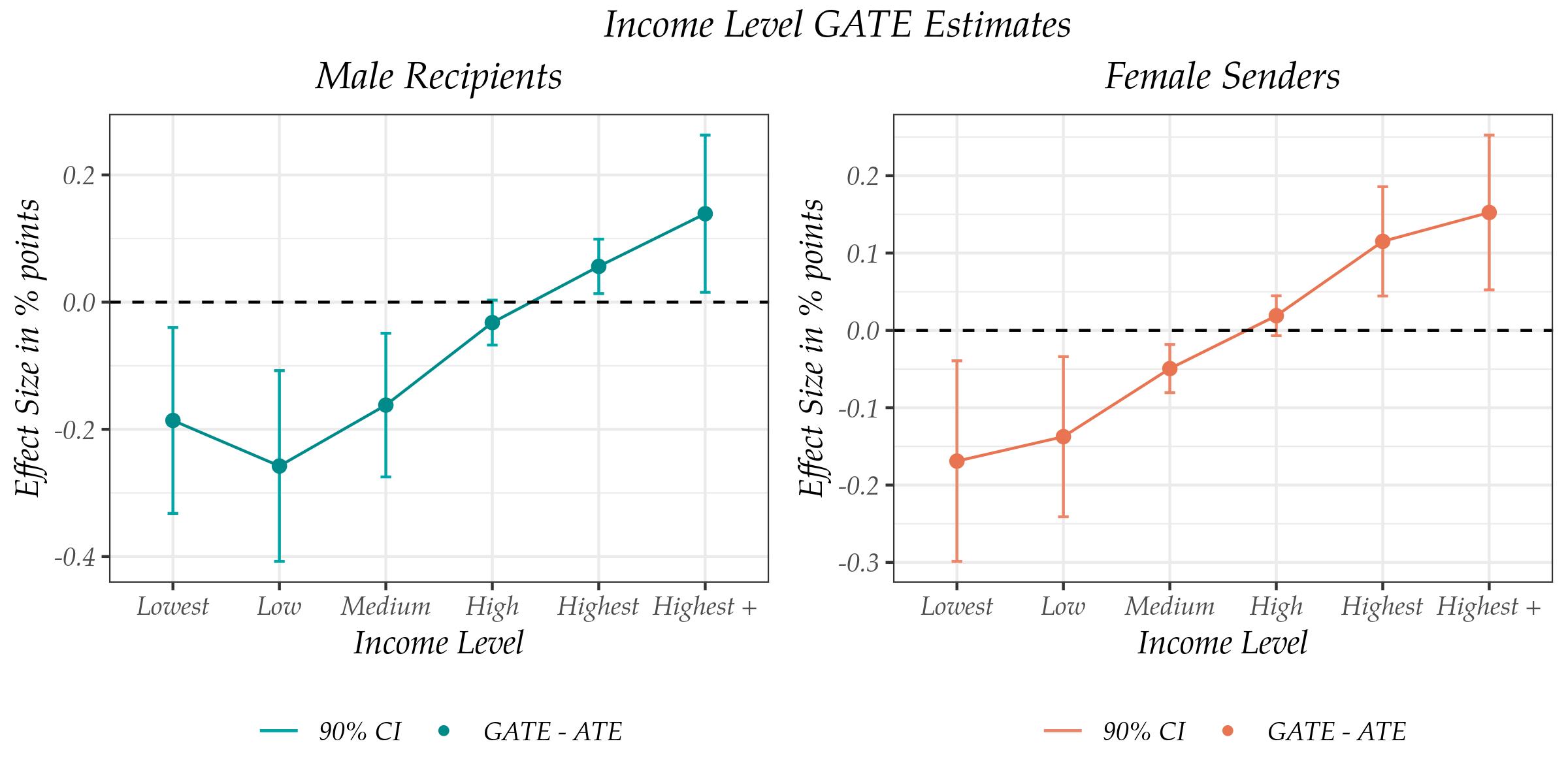}
    \caption*{\footnotesize \textit{Note:} Effects in $\%$ points as GATE deviations from the ATE (zero dotted line) with $90\%$ confidence intervals.}
\end{figure}

The results of the Wald test suggest heterogeneous effects with regard to the income level for males, both for the recipient as well as the corresponding sender, however, no evidence of heterogeneity in case of females. Furthermore, the heterogeneous effects for males are statistically different from the average effect as well, indicating an explicit pattern, while none of this is the case for females. The respective income level GATE estimates are depicted in Figure \ref{fig:male_gates} for the recipient and the sender in the male sample. The corresponding results for the female sample are presented in Figure \ref{fig:female_gates} in Appendix \ref{App:D}.

Concerning the male sample, we observe a clear increasing trend of the GATEs for increasing levels of income. As such, for a male recipient, the effect of weekly sport activity in contrast to no sport is greater, the higher the income level of the male recipient himself, and the higher the income level of the female sender, too. As a result, male users with a higher income level, benefit from a regular sport activity on a weekly basis in comparison to no sport, more than male users with a lower income level. This implies that particularly the wealthy males, who earn more than 100'000 EUR in a year, can increase their contact chances the most by practicing sport on a weekly basis. In a similar vein, male users having a potential female partner with high income level, benefit from the higher sport frequency more than the male users, which have a potential female partner with low income level. This pattern suggests also that more wealthy female users value the regular sport activity of a male user more. In addition, not only are these heterogeneous effects statistically relevant, the substantive relevance is documented, too, as the effect sizes are relatively large. As such, the magnitudes of the income level GATEs are ranging from $1.06\%$ points to $1.46\%$ points with respect to the income level of a male recipient, and similarly, from $1.15\%$ points to $1.47\%$ points with respect to the income level of a female sender, in reference to the average effect of $1.32\%$ points. This implies an increase in the message probability of at least $42.40\%$ for the low income users, up to an increase of $58.80\%$ for the high income users, respectively. This results in a $16.40\%$ difference in message probability solely due to the user's income. A simple back of the envelope calculation reveals this difference in income levels to amount to 4 extra messages out of 1000 profile visits.

\vspace{-0.35cm}

As opposed to the male sample, we do not find such evidence of heterogeneity, if we switch the roles of the recipient and the sender (see Figure \ref{fig:female_gates} in Appendix \ref{App:D}). As such, even though we observe a similar increasing pattern for female recipients associated with male senders, the estimated effects lack statistical relevance. 

\begin{figure}[H]
    \centering
    \caption{Heterogeneous Effects of Sport Activity based on Sport for Females}
    \label{fig:female_gates_sport}
    \includegraphics[width=0.95\textwidth]{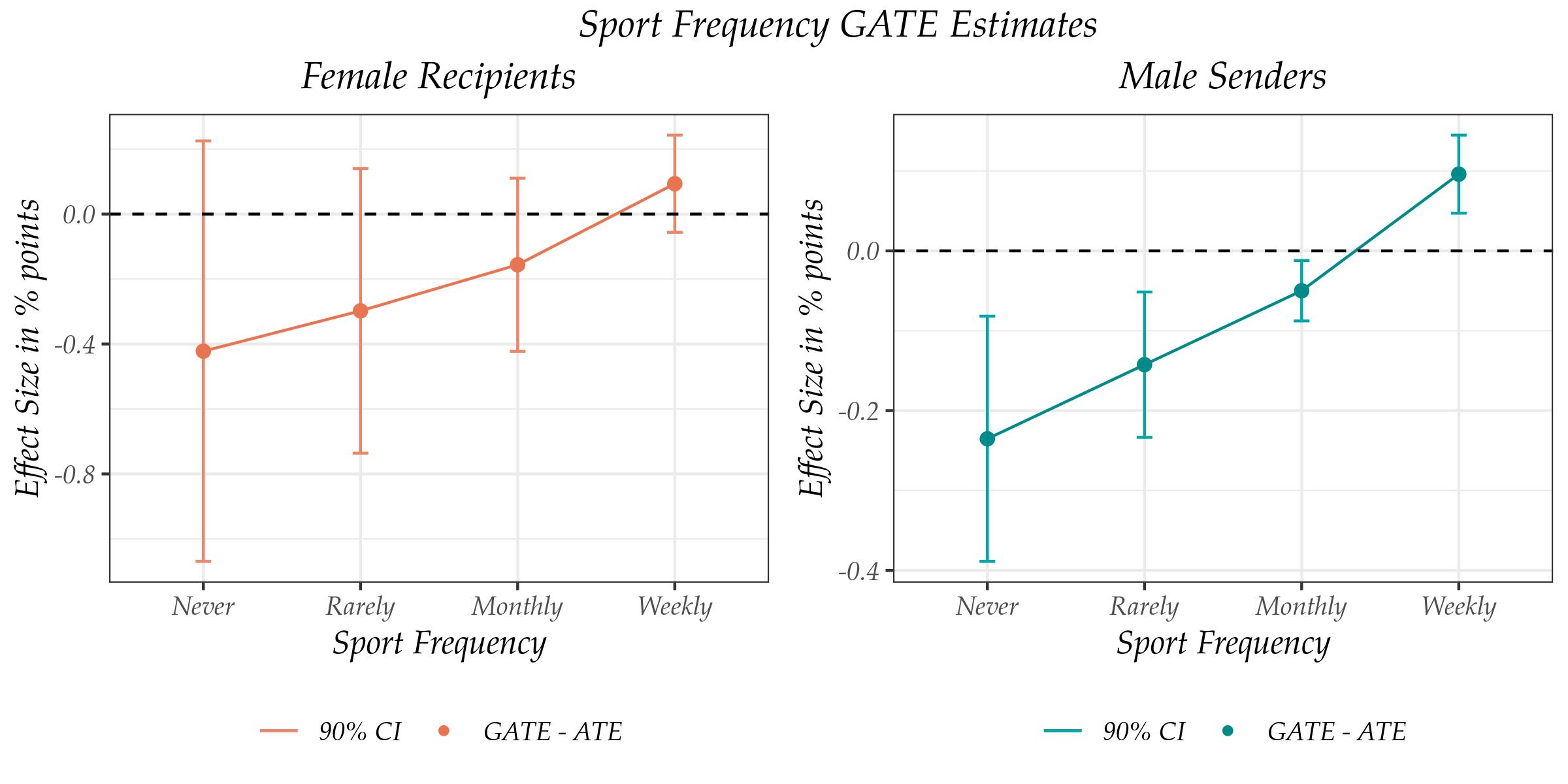}
    \caption*{\footnotesize \textit{Note:} Effects in $\%$ points as GATE deviations from the ATE (zero dotted line) with $90\%$ confidence intervals.}
\end{figure}

However, in contrast to the results for income heterogeneity, we find supportive evidence for heterogeneity for females in terms of the sport activity, while no such evidence is detected for males. As such, for females, both the Wald test of effect equality as well as the $t$-tests for differences from the average effect suggest presence of heterogeneity with respect to the level of sport frequency of the male sender with a clear increasing pattern, whilst the heterogeneity with respect to the female recipient lacks the statistical precision. Contrarily, for the male sample, even though we observe a similar increasing pattern as for the female sample, the statistical relevance is, however, absent. The corresponding results for the female sample regarding the sport frequency GATE estimates are presented in Figure \ref{fig:female_gates_sport} for both the female recipient and the male sender. The respective results for the male sample are depicted in Figure \ref{fig:male_gates_sport} in Appendix \ref{App:D}.

The heterogeneity results with respect to the sport frequency suggest that for a female user, the effect of a weekly sport activity in contrast to no sport is greater, the higher the sport frequency of the potential male partner. Thus, females enjoy a higher effect of their own weekly sport activity, if the sport activity of a potential male partner is on a weekly basis as well. This further suggests that sporty male users appreciate sporty female users more. Nevertheless, despite the clear statistical pattern of the heterogeneity itself, in this case the overall substantive implications remain rather limited as the effect sizes are only moderate, ranging from $0.08\%$ points to $0.41\%$ points, given the average effect of $0.31\%$ points. Additionally, neither for the average effect nor for the respective group effects the presence of an actual null effect can be ruled out.

\vspace{-0.35cm}

Further results of the Wald tests regarding the remaining heterogeneity variables do not indicate differential effects at conventional significance levels in terms of age or the mutual user distance, concerning both males as well as females. Neither do the differences of the estimated GATEs from the ATE support the evidence for heterogeneous effects. Furthermore, although the Wald test of equality of GATEs based on the education level suggests presence of effect heterogeneity, the differences from the average effect are not statistically relevant and lack an explicit pattern.\footnote{The exhaustive results for the effect heterogeneity analysis can be found in Appendix \ref{App:D}.}

Altogether, based on the GATEs analysis, we conclude to find a supporting evidence, both statistical as well as substantive, for heterogeneity in terms of the income level for males and statistical, however, not substantive evidence, in terms of the sport frequency for females, whereas, we find lack of evidence in general, for heterogeneous effects along the age, distance and education level for both males and females.

\begin{figure}[H]
    \centering
     \caption{Distribution of the Individualized Effects of Sport Activity for Males and Females}
    \label{fig:iates}
    \includegraphics[width=0.95\textwidth]{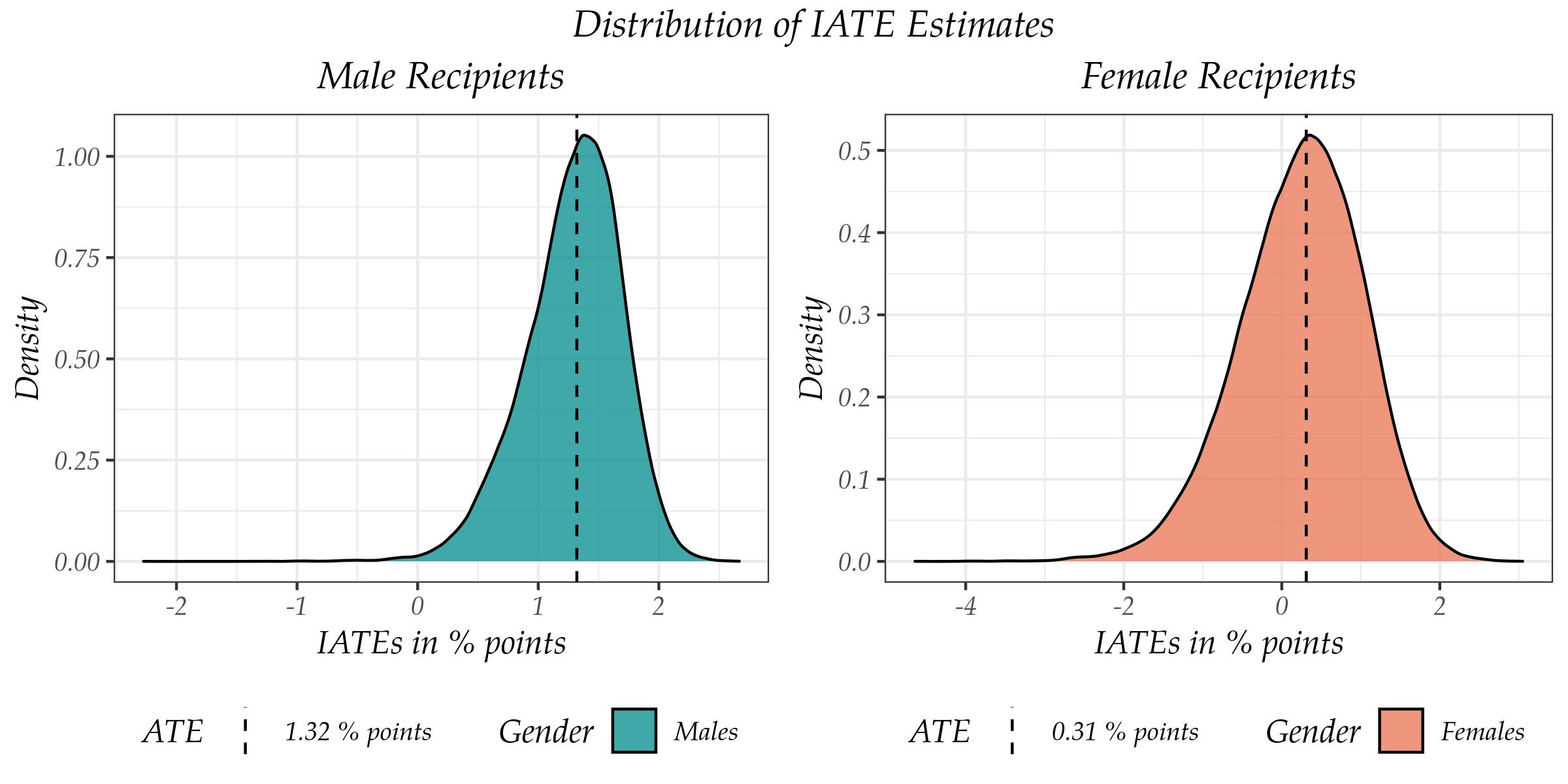}
   \caption*{\footnotesize \textit{Note:} Distribution of IATEs smoothed with the Epanechnikov kernel using the Silverman's bandwidth.}
\end{figure}

Additionally, in order to gain more insight for the effect heterogeneity, we analyze the effects on the finest level possible and study the underlying individualized average treatment effects. Figure \ref{fig:iates} provides the distribution of the IATEs for the weekly vs. never comparison, for both the male as well as the female sample, respectively. In both cases, we observe that there is indeed substantial heterogeneity in the considered effects as the effect distributions are noticeably spread out around the mean, i.e. the realized ATE.\footnote{Part of the observed variability is also due to estimation uncertainty: the average standard error for the IATEs is $0.61$ for the male and $1.06$ for the female sample, respectively.} Additionally, we see that for males, virtually all effects are positive, while for females, about half of the effects are positive and half are negative. This further substantiates the findings on the aggregated levels in terms of the GATEs and the ATE.

\vspace{-0.35cm}

To understand these effect distributions more thoroughly, we apply the $k$-means\plusplus{} clustering \parencite{arthur2006} to provide further descriptive evidence of the dependence of the effects on the heterogeneity variables \parencite[see][for an analogous approach]{cockx2020}. For this purpose, we perform the clustering by using the IATEs for the weekly vs. never comparison to form distinct clusters, which we sort increasingly according to the mean effect size. We then describe the clusters by the means of the corresponding heterogeneity variables, which however, have not been used to form the clusters. Table \ref{tab:kmeans} presents the clusters for the IATEs of the male and female sample, respectively.

In general, the clustering reveals consistent patterns with the heterogeneity analysis based on the GATEs. For the male sample, the increasing effects of the sport frequency along the clusters are associated with an increasing level of income on both the recipient as well as the sender side. As such, the lowest effects of sport are clearly for the users with the lowest income level, and vice-versa, the highest effects are evidently for those users with the highest income level. Complementary to the GATEs analysis, the clustering additionally reveals similar increasing patterns in terms of education level and the sport frequency for males. This indicates a further positive relationship, which, however, lacks statistical relevance within the GATEs analysis. Nonetheless, the clustering does not find any particularly clear patterns in terms of age or mutual distance, which is consistent with the GATE estimates. 

\pagebreak

In case of the female sample, complementary to the GATEs, the clusters suggest an increasing effect for higher sport frequency as documented within the GATEs analysis. However, according to the cluster analysis, this holds true not only for the sender, but also for the recipient side, for which the statistical evidence in terms of the GATEs is missing. In a similar vein, the clusters also suggest a relevant heterogeneity with respect to the income level with an increasing pattern. Furthermore, as for the male clusters, also the female clusters suggest additionally a positive association of the IATEs with the education level, however, no apparent indication of heterogeneity for age or mutual distance.

\begin{table}[H]
\centering
\caption{Clusters of the Individualized Effects of Sport Activity for Males and Females}\label{tab:kmeans}
\scriptsize
\resizebox{1.0\textwidth}{!}{%
\begin{tabular}{lrrrrrrrrrrr}
 \toprule
& \multicolumn{5}{c}{\cellcolor{white}Males} & & \multicolumn{5}{c}{\cellcolor{white}Females}\\
 \midrule
\cellcolor{white}\textit{Clusters} & 1 & 2 & 3 & 4 & 5 & & 1 & 2 & 3 & 4 & 5 \\ 
   \midrule
IATEs: Weekly vs. Never & \cellcolor{malegreen!15}0.41 & \cellcolor{malegreen!25}0.88 & \cellcolor{malegreen!35}1.22 & \cellcolor{malegreen!55}1.52 & \cellcolor{malegreen!75}1.85 & & \cellcolor{femalegreen!25}-1.41 & \cellcolor{femalegreen!35}-0.52 & \cellcolor{femalegreen!55}0.12 & \cellcolor{femalegreen!75}0.71 & \cellcolor{femalegreen!95}1.38 \\ 
   \midrule
\textit{Recipient Features} &  &  &  &  &  &  &  &  &  &  &  \\ 
   \midrule
Age & 45.77 & 45.19 & 44.57 & 43.80 & 43.71 & & 33.53 & 37.07 & 38.39 & 38.15 & 36.90 \\ 
  Education Level & 3.51 & 3.86 & 4.24 & 4.58 & 4.76 & & 3.65 & 3.74 & 3.80 & 3.93 & 4.05 \\ 
  Income Level & \cellcolor{malegreen!15}3.69 & \cellcolor{malegreen!25}3.98 & \cellcolor{malegreen!35}4.26 & \cellcolor{malegreen!55}4.57 & \cellcolor{malegreen!75}4.83 & & 2.98 & 3.16 & 3.32 & 3.46 & 3.53 \\ 
  Sport Frequency & 1.85 & 2.16 & 2.37 & 2.46 & 2.49 & & 1.55 & 1.92 & 2.15 & 2.34 & 2.43 \\ 
   \midrule
\textit{Sender Features} &  &  &  &  &  &  &  &  &  &  &  \\ 
   \midrule
Age & 43.94 & 43.09 & 42.37 & 41.53 & 41.43 & & 36.33 & 40.15 & 41.68 & 41.46 & 40.23 \\ 
  Education Level & 3.46 & 3.69 & 3.96 & 4.16 & 4.34 & & 3.74 & 3.82 & 3.93 & 4.06 & 4.25 \\ 
  Income Level & \cellcolor{malegreen!15}2.94 & \cellcolor{malegreen!25}3.25 & \cellcolor{malegreen!35}3.52 & \cellcolor{malegreen!55}3.79 & \cellcolor{malegreen!75}4.07 & & 3.51 & 3.82 & 3.99 & 4.10 & 4.14 \\ 
  Sport Frequency & 1.86 & 2.03 & 2.16 & 2.24 & 2.32 & & \cellcolor{femalegreen!25}1.91 & \cellcolor{femalegreen!35}2.06 & \cellcolor{femalegreen!55}2.15 & \cellcolor{femalegreen!75}2.27 & \cellcolor{femalegreen!95}2.44 \\
   \midrule
\textit{Shared Features} &  &  &  &  &  &  &  &  &  &  &  \\ 
   \midrule
Distance & 70.39 & 71.65 & 70.23 & 70.16 & 64.91 & & 64.43 & 66.22 & 67.10 & 66.36 & 62.97 \\ 
   \midrule
\textit{Observations} &  &  &  &  &  &  &  &  &  &  &  \\ 
   \midrule
Share & 0.07 & 0.18 & 0.29 & 0.31 & 0.15 & & 0.07 & 0.20 & 0.29 & 0.29 & 0.15 \\ 
  Total & 2288 & 6439 & 10153 & 10826 & 5252 & & 3753 & 11048 & 15896 & 15484 & 8047 \\ 
   \bottomrule
   \multicolumn{12}{l}{\scriptsize \textit{Note:} Means of clustered effects sorted in an increasing order, matched with the heterogeneity variables. Color}\\
   \multicolumn{12}{l}{\scriptsize intensity represents the corresponding effect sizes and highlights the relevant GATEs.}
   \end{tabular}%
   }
\end{table}

\vspace{-0.2cm}

In addition to the GATE analysis, the clusters further allow for a more detailed description of the IATEs based on the user characteristics, beyond the pre-specified subset of heterogeneity variables. Notably, the cluster analysis reveals a particular relationship between the IATEs and the behavior and preferences of the users, both for males and females. As such, higher IATEs are associated with increasing preference to find the significant other and to have an intimate relationship, as well as with increasing satisfaction of own appearance. In contrast, lower IATEs are associated with increasing smoking frequency, as well as increasing preference for media consumption and comfortable dining. These insights provide not only a better understanding of the specific individualized effects of sport activity on the contact chances, but might serve as a basis and guidance for a selection of relevant heterogeneity variables in future research. An overview of the relevant clusters with variable description is provided in Table \ref{tab:clusters} in Appendix \ref{App:D}.

Overall, the cluster analysis of the IATEs emphasizes the results from the GATEs, and as such provides additional evidence for the income heterogeneity for males, as well as the sport heterogeneity for females. Moreover, it reveals further descriptive evidence for increasing effects based on education level, albeit no particular heterogeneity patterns for age or mutual distance. Lastly, it provides valuable insights for additional heterogeneity channels such as relationship preferences.

\subsection{Placebo Test}\label{sec:placebo}
In our analysis of the effect of sport activity on contact chances, we assume that the treatment, i.e. the sport frequency is observed once a profile of a recipient has been visited by a sender. However, as discussed in Section \ref{sec:identification}, the sport frequency might potentially be entailed in the matching score, which is observable already before the actual profile visit as part of the shortlist of potential partners suggested by the online dating platform. If that would be the case, the sport frequency could potentially indirectly influence already the decision to visit the profile, and not only the decision to send a text message after a profile visit. However, even under such circumstances, this would not violate the CIA \textit{per se}, but rather reduce the size of the estimated effect, which could be then interpreted as a lower bound of the true underlying effect. In order to examine if such mechanism takes place in our setting, we conduct a placebo test inspired by \textcite{imbens2009} to assess the validity of the CIA by testing for a zero effect on an outcome variable assumed to be unaffected by the treatment, here the decision to visit the user profile. Accordingly, we redo our main analysis, while swapping the message outcome for a visit outcome. Thus, we estimate the average treatment effects of sport frequency on the visit probability, given the same conditioning set. Therefore, if the sport frequency is, as assumed, not part of the matching score, its effect on the probability to visit a user profile should be equal to zero.

In order to implement such placebo test, we first need to impute the 'potential' visits, as by construction, we only observe the realized visits. For a given user, we consider all registered user profiles with opposite sex and within a specified distance radius as potential visits.\footnote{We restrict the potential visits to opposite sex as we observe only heterosexual users in our sample. Furthermore, we restrict the distance of potential users due to dimensionality concerns, as the share of the realized visits would otherwise be almost completely diminished, if unrestricted. Here, we remain rather conservative and set the potential distance to 95\% of the maximum observed distance of an actual realized visit.} We end up with a sample consisting of 38'552'821 observations, out of which 178'372 represent the actual realized and the rest the imputed potential visits. Analogously as in the main analysis, we split the sample into a male and a female sample. Furthermore, due to the computational feasibility and general consistency of the analysis, we randomly draw an identically sized male and female sample as in the main estimation, such that we replicate the corresponding sport frequency shares, too.\footnote{We repeated the random draw several times, while the results remained qualitatively robust.} A similar approach to impute the potential visits has been used also in previous studies focusing on online dating platforms \parencite{bruch2016}. 

\begin{table}[H]
\centering
\caption{Average Effects of Sport Activity on the Visit Chances for Males and Females}\label{tab:placebo}
\resizebox{1.0\textwidth}{!}{%
\begin{tabular}{lp{1.1cm}p{1.1cm}p{1.1cm}p{1.1cm}p{1.1cm}p{1.1cm}p{1.1cm}p{1.1cm}p{1.1cm}}
  \toprule
 & \multicolumn{4}{c}{Males} & & \multicolumn{4}{c}{Females} \\
  \midrule
 &  \cellcolor{white}\textit{Never} &  \cellcolor{white}\textit{Rarely} &  \cellcolor{white}\textit{Monthly} &  \cellcolor{white}\textit{Weekly} &  & \cellcolor{white}\textit{Never} &  \cellcolor{white}\textit{Rarely} &  \cellcolor{white}\textit{Monthly} & \cellcolor{white}\textit{Weekly} \\ 
  \midrule
\cellcolor{white}\textit{Never} & \cellcolor{malegreen!15}0.23 &  &  &  &  & \cellcolor{femalegreen!95}0.66 &  &  &  \\ 
   & \cellcolor{malegreen!15}\scalebox{0.75}{(0.14)} &  &  &  &  & \cellcolor{femalegreen!95}\scalebox{0.75}{(0.15)} &  &  &  \\ 
\cellcolor{white}\textit{Rarely} & 0.29 & \cellcolor{malegreen!75}0.52 &  &  &  & -0.07 & \cellcolor{femalegreen!75}0.59 &  &  \\ 
   & \scalebox{0.75}{(0.24)} & \cellcolor{malegreen!75}\scalebox{0.75}{(0.20)} &  &  &  & \scalebox{0.75}{(0.22)} & \cellcolor{femalegreen!75}\scalebox{0.75}{(0.16)} &  &  \\ 
\cellcolor{white}\textit{Monthly} & 0.08 & -0.22 & \cellcolor{malegreen!35}0.31 &  &  & -0.20 & -0.13 & \cellcolor{femalegreen!35}0.46 &  \\ 
   & \scalebox{0.75}{(0.16)} & \scalebox{0.75}{(0.22)} & \cellcolor{malegreen!35}\scalebox{0.75}{(0.08)} &  &  & \scalebox{0.75}{(0.17)} & \scalebox{0.75}{(0.17)} & \cellcolor{femalegreen!35}\scalebox{0.75}{(0.08)} &  \\ 
\cellcolor{white}\textit{Weekly} & \cellcolor{white}0.18 & \cellcolor{white}-0.12 & 0.10 & \cellcolor{malegreen!55}0.41 &  & -0.08 & \cellcolor{white}-0.01 & 0.12 & \cellcolor{femalegreen!55}0.58 \\ 
  & \cellcolor{white}\scalebox{0.75}{(0.15)} & \cellcolor{white}\scalebox{0.75}{(0.21)} & \scalebox{0.75}{(0.10)} & \cellcolor{malegreen!55}\scalebox{0.75}{(0.06)} &  & \scalebox{0.75}{(0.16)} & \cellcolor{white}\scalebox{0.75}{(0.17)} & \scalebox{0.75}{(0.10)} & \cellcolor{femalegreen!55}\scalebox{0.75}{(0.07)} \\ 
   \bottomrule
   \multicolumn{10}{l}{\footnotesize \textit{Note:} Effects in $\%$ points. Potential outcomes on the diagonal. Standard errors in parentheses. Significance levels}\\
   \multicolumn{10}{l}{\footnotesize refer to: $^{***}<0.01$, $^{**}<0.05$, $^{*}<0.1$. Color intensity represents the corresponding level sizes.}
\end{tabular}%
}
\end{table}

\vspace{-0.2cm}

Table \ref{tab:placebo} summarizes the ATE results of the Modified Causal Forest estimation for the placebo test. First of all, we observe that the potential outcomes for both males and females do not exhibit any particular upward or downward trend as is the case for the main analysis. Furthermore, for neither the male nor the female sample, we find evidence for statistically relevant effects. Moreover, the effect sizes and the levels of potential outcomes are an order of magnitude lower than our main results, being effectively zero in terms of the substantive relevance. Even though the results of such placebo tests do not completely rule out the possibility of a presence of an effect on the visit probability, they provide a supportive evidence that this is, indeed, not the case. Hence, we conclude that our main analysis estimates the full causal effects of sport activity on the contact chances, rather than only lower bounds thereof.

\section{Discussion}\label{sec:discussion}

The main objective of this paper was to analyze the effect of sport activity on human mating. Following this objective, we examined the effect of sport frequency on contact chances based on a unique dataset from an online dating platform and applying the Modified Causal Forest estimator \parencite{lechner2018}. We found that for male users, doing sport on a weekly basis increases the probability to receive a first message by more than 50\% relatively to not doing sport at all, while for female users, we do not find evidence for such an effect. In addition, we uncover important effect heterogeneities. In particular, the effect of sport frequency on contact chances increases with higher income for male, but not for female, users.

This paper offers notable implications for research and practice. First, this study contributes to the literature on human mating. In particular, we demonstrate that sport activity, as an essential  behavioral trait and pivotal information on online dating platforms, enfolds a causal effect on contact chances. In turn, this paper overcomes limitations of previous work that did not consider or comprehensively map the effect of sport activity on human mating. Moreover, this paper expands previous work on the effects of sport activity by demonstrating that sport activity does not only affect physical/mental health and social and economic conditions, as well-documented by prior research \parencites{strong2005}{lechner2009}, but also one of the most decisive spheres of human existence, that is human mating.

Second, this paper advances empirical approaches for assessing causal effects in large-dimensional data environments, as applicable, for example, to remote-sensing data.  In particular, this research applies a very flexible estimation procedure, which offers not only greater flexibility in considering (interrelated effects of) covariates, but also a systematic analysis of the underlying heterogeneities of the effects on different levels of aggregation \parencite{lechner2018}. Thus, this paper may support future research in analyzing human behavior in large-dimensional data environments. However, even though the causal machine learning approach is capable of detecting statistically relevant heterogeneities, it is crucial to assess also its substantive relevance. Following this notion, the effect heterogeneities in this research provide different perspectives on practical implications. In particular, the increasing effect of sport activity on contact chances with higher income for male users is both statistically justified as well as substantially relevant, leading to the above mentioned implication. Contrarily, potential implications, resulting from the observation that the effect of sport frequency on contact chances increases with higher sport frequency for females, are limited as the particular evidence in our setting is not substantially relevant, even though it is statistical justified. In addition to the main heterogeneity analysis, the post-estimation descriptive cluster analysis of the most disaggregated effects provides additional insights for possible heterogeneity channels, such as the education level or relationship preferences of the users.

Third, this study may support individuals to increase their chances of finding a mate on online dating platforms by demonstrating if and to what extent sport activity contributes to the likelihood to be recognized. In particular, men may benefit from the insights of this research by being aware that sport on a weekly basis relative to no sport can increase their probability to receive a first message by more than 50\%, or even up to 60\% in case of higher income individuals, while for women the effect of sport activity on the contact chances is not entirely evident. Thus, this study may incentivise individuals to increase the level of sport activity, not only because of the well-documented effects on, for example, health \parencite{penedo2005}, but also for their chances of finding a mate.

Moreover, from a public health perspective, this paper provides empirical reasoning for justifying and evaluating incentives for public health promotion due to the impact of sport activities for human partnering, family planning, and reproduction. 

Finally, this paper may serve practitioners, namely product developers and software engineers, as a foundation to improve the architecture of online dating platforms, including interface designs and matching algorithms. In particular, this study points out the relevance of sport activity for mate evaluation and selection patterns, while considering effect heterogeneities based on established socio-demographic characteristics at the same time. In turn, this research may help practitioners to assess humans’ mate evaluation and selection in much more detail and, correspondingly, to evaluate improvements of the architecture of online dating platforms (e.g., customized weighting of sport activity in matching algorithms or specific placement of information on sport activity on individual profile pages). In a similar vein, the insights of this research are applicable to engineer architectures of other platforms with a likewise high degree of interpersonal computer-mediated interaction, for example, social networks.

\pagebreak
\printbibliography
\pagebreak
\appendix 

\section{Descriptive Statistics}\label{App:B}

\begin{table}[H]
\centering
\caption{Descriptive Statistics for the User Sample}
\footnotesize
\begin{tabular}{rrrrr}
\toprule
  & \cellcolor{white}Mean & \cellcolor{white}SD & \cellcolor{white}Min & \cellcolor{white}Max \\ 
  \midrule
   \cellcolor{white}\textit{Sport Frequency} & & & & \\ 
\midrule
Never & 0.13 & 0.34 & 0.00 & 1.00 \\ 
  Rarely & 0.10 & 0.31 & 0.00 & 1.00 \\ 
  Monthly & 0.29 & 0.45 & 0.00 & 1.00 \\ 
  Weekly & 0.48 & 0.50 & 0.00 & 1.00 \\ 
  \midrule
  \cellcolor{white}\textit{Demographic Features} & & & & \\ 
\midrule
 Gender (=1 if female) & 0.48 & 0.50 & 0.00 & 1.00 \\ 
  Age (in years) & 40.05 & 11.41 & 18.00 & 82.00 \\ 
   \midrule
   \cellcolor{white}\textit{Income Level}  & & & & \\ 
\midrule
Lowest & 0.12 & 0.32 & 0.00 & 1.00 \\ 
  Low & 0.16 & 0.37 & 0.00 & 1.00 \\ 
  Medium & 0.20 & 0.40 & 0.00 & 1.00 \\ 
  High & 0.24 & 0.43 & 0.00 & 1.00 \\ 
  Highest & 0.22 & 0.42 & 0.00 & 1.00 \\ 
  Highest+ & 0.06 & 0.24 & 0.00 & 1.00 \\ 
   \midrule
   \cellcolor{white}\textit{Education Level}  & & & & \\ 
\midrule
Lowest & 0.00 & 0.06 & 0.00 & 1.00 \\ 
  Low & 0.08 & 0.27 & 0.00 & 1.00 \\ 
  Medium & 0.37 & 0.48 & 0.00 & 1.00 \\ 
  High & 0.16 & 0.37 & 0.00 & 1.00 \\ 
  Highest & 0.39 & 0.49 & 0.00 & 1.00 \\ 
  \bottomrule
  \multicolumn{5}{l}{\footnotesize \textit{Note:} Main variables describing the population displayed.}
  \end{tabular}
\end{table}

\begin{figure}[H]
    \centering
    \includegraphics[width=0.99\textwidth]{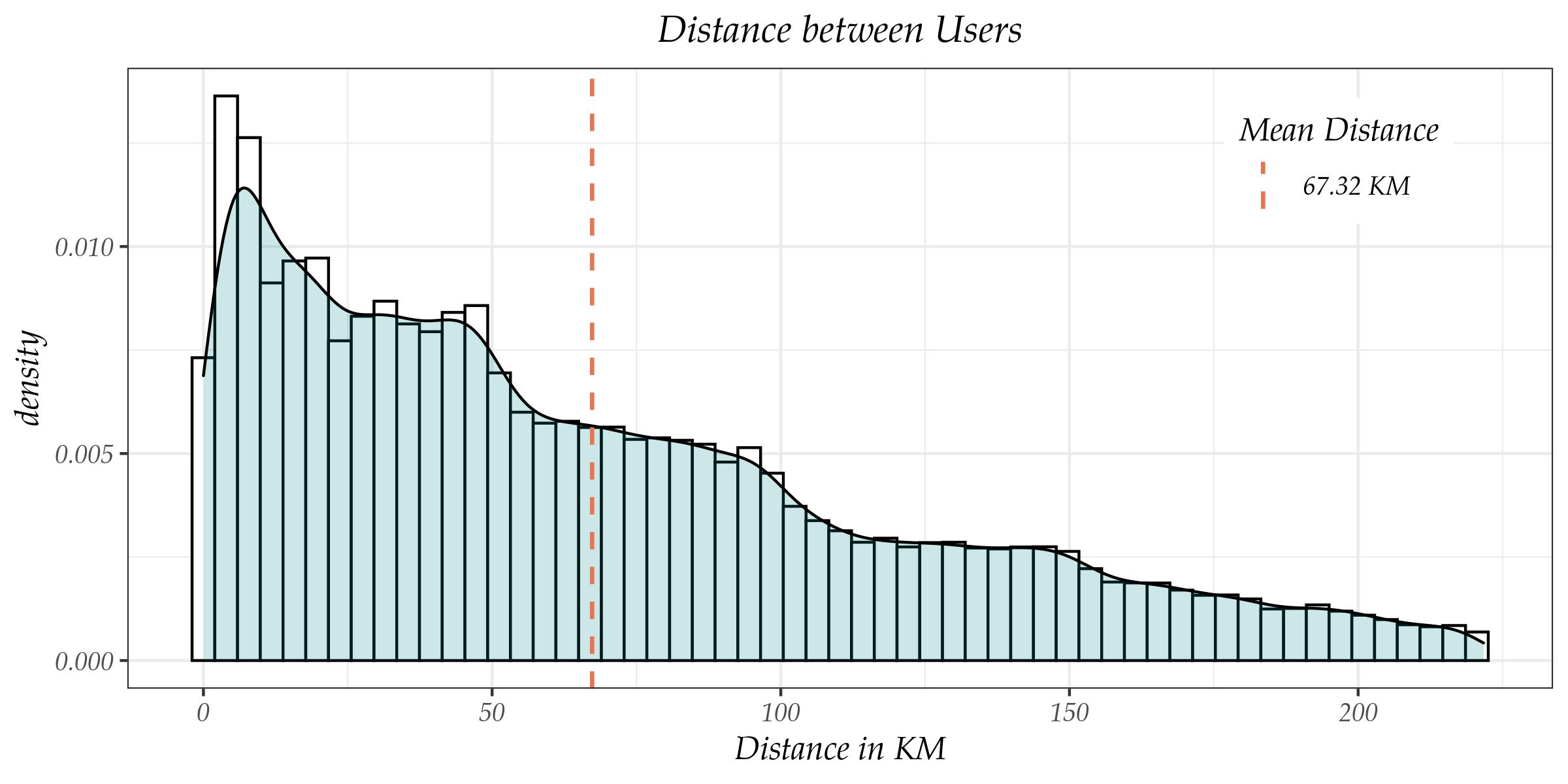}
    \caption{Distribution of the Distance between Users}
    \label{fig:distance}
\end{figure}

\begin{table}[H]
\centering
\footnotesize
\caption{Descriptive Statistics by Sport Frequency for Female Sample} 
\begin{tabular}{lrrrrr}
  \toprule
 & \cellcolor{white}Never & \cellcolor{white}Rarely & \cellcolor{white}Monthly & \cellcolor{white}Weekly & \cellcolor{white}Total \\ 
  \midrule
  \cellcolor{white}\textit{Outcome} &  & & & & \\ 
  \midrule
First Message & 0.11 & 0.10 & 0.10 & 0.11 & 0.11 \\ 
\midrule
\cellcolor{white}\textit{Recipient Features} &  & & & & \\ 
\midrule
   Age & 35.49 & 37.65 & 37.84 & 37.83 & 37.53 \\ 
   &  & & & & \\ 
   \textit{Income Level} &  & & & & \\ 
   Lowest & 0.20 & 0.20 & 0.10 & 0.07 & 0.10 \\ 
   Low & 0.28 & 0.23 & 0.19 & 0.14 & 0.18 \\ 
   Medium & 0.28 & 0.25 & 0.23 & 0.22 & 0.23 \\ 
   High & 0.17 & 0.21 & 0.27 & 0.33 & 0.28 \\ 
   Highest & 0.05 & 0.11 & 0.18 & 0.21 & 0.18 \\ 
   Highest+ & 0.01 & 0.01 & 0.03 & 0.04 & 0.03 \\ 
   &  & & & & \\ 
  \textit{Education Level} &  & & & & \\ 
   Lowest & 0.00 & 0.00 & 0.00 & 0.00 & 0.00 \\ 
   Low & 0.08 & 0.06 & 0.02 & 0.02 & 0.03 \\ 
   Medium & 0.61 & 0.52 & 0.41 & 0.36 & 0.42 \\ 
   High & 0.20 & 0.21 & 0.21 & 0.22 & 0.21 \\ 
   Highest & 0.11 & 0.21 & 0.35 & 0.41 & 0.33 \\
   &  & & & & \\ 
  \midrule
  \cellcolor{white}\textit{Sender Features} &  & & & & \\ 
\midrule
  Age & 38.57 & 40.80 & 41.11 & 41.07 & 40.75 \\
  &  & & & & \\ 
  \textit{Income Level} &  & & & & \\ 
  Lowest & 0.09 & 0.07 & 0.05 & 0.04 & 0.05 \\ 
  Low & 0.16 & 0.13 & 0.09 & 0.07 & 0.09 \\ 
  Medium & 0.23 & 0.21 & 0.17 & 0.16 & 0.18 \\ 
  High & 0.26 & 0.26 & 0.27 & 0.27 & 0.27 \\ 
  Highest & 0.21 & 0.26 & 0.32 & 0.35 & 0.32 \\ 
  Highest+ & 0.04 & 0.07 & 0.10 & 0.11 & 0.09 \\ 
  &  & & & & \\ 
  \textit{Education Level} &  & & & & \\
  Lowest & 0.00 & 0.00 & 0.00 & 0.00 & 0.00 \\ 
  Low & 0.11 & 0.10 & 0.07 & 0.05 & 0.06 \\ 
  Medium & 0.47 & 0.41 & 0.32 & 0.28 & 0.33 \\ 
  High & 0.14 & 0.14 & 0.16 & 0.17 & 0.16 \\ 
  Highest & 0.28 & 0.35 & 0.45 & 0.50 & 0.44 \\ 
  &  & & & & \\ 
  \textit{Sport Frequency} &  & & & & \\ 
  Never & 0.16 & 0.14 & 0.11 & 0.09 & 0.11 \\ 
  Rarely & 0.13 & 0.12 & 0.10 & 0.09 & 0.10 \\ 
  Monthly & 0.29 & 0.30 & 0.31 & 0.28 & 0.29 \\ 
  Weekly & 0.42 & 0.44 & 0.49 & 0.55 & 0.50 \\ 
  &  & & & & \\ 
  \midrule
  \cellcolor{white}\textit{Observations} &  & & & & \\ 
\midrule
  Total Share & 0.12 & 0.09 & 0.29 & 0.49 & 1.00 \\ 
  Total Observations & 13'408 & 98'33 & 31'801 & 53'414 & 108'456 \\ 
   \bottomrule
   \multicolumn{6}{l}{\footnotesize \textit{Note:} Means of variables displayed in all columns.}
\end{tabular}
\end{table}

\begin{table}[H]
\centering
\footnotesize
\caption{Descriptive Statistics by Sport Frequency for Male Sample} 
\begin{tabular}{lrrrrr}
  \toprule
 & \cellcolor{white}Never & \cellcolor{white}Rarely & \cellcolor{white}Monthly & \cellcolor{white}Weekly & \cellcolor{white}Total \\ 
  \midrule
\cellcolor{white}\textit{Outcome} &  & & & & \\ 
\midrule
First Message & 0.02 & 0.03 & 0.03 & 0.04 & 0.04 \\ 
\midrule
\cellcolor{white}\textit{Recipient Features} &  & & & & \\ 
\midrule
   Age & 44.88 & 46.21 & 45.56 & 43.43 & 44.37 \\ 
  &  & & & & \\ 
  \textit{Income Level} &  & & & & \\ 
   Lowest & 0.06 & 0.03 & 0.02 & 0.02 & 0.02 \\ 
   Low & 0.15 & 0.08 & 0.05 & 0.04 & 0.05 \\ 
   Medium & 0.21 & 0.19 & 0.14 & 0.12 & 0.14 \\ 
   High  & 0.22 & 0.32 & 0.27 & 0.25 & 0.26 \\ 
   Highest  & 0.30 & 0.30 & 0.41 & 0.42 & 0.40 \\ 
   Highest+ & 0.05 & 0.07 & 0.11 & 0.15 & 0.13 \\ 
  &  & & & & \\ 
 \textit{ Education Level} &  & & & & \\ 
   Lowest & 0.00 & 0.01 & 0.00 & 0.00 & 0.00 \\ 
   Low & 0.11 & 0.07 & 0.03 & 0.02 & 0.03 \\ 
   Medium & 0.37 & 0.36 & 0.24 & 0.17 & 0.22 \\ 
   High & 0.15 & 0.20 & 0.15 & 0.15 & 0.16 \\ 
   Highest & 0.37 & 0.37 & 0.58 & 0.66 & 0.59 \\ 
  &  & & & & \\ 
  \midrule
  \cellcolor{white}\textit{Sender Features} &  & & & & \\ 
\midrule
  Age & 43.15 & 44.20 & 43.37 & 41.19 & 42.19 \\ 
  &  & & & & \\ 
 \textit{Income Level} &  & & & & \\ 
  Lowest & 0.13 & 0.09 & 0.07 & 0.06 & 0.07 \\ 
  Low  & 0.20 & 0.18 & 0.13 & 0.11 & 0.13 \\ 
  Medium & 0.25 & 0.25 & 0.23 & 0.22 & 0.23 \\ 
  High  & 0.26 & 0.29 & 0.31 & 0.33 & 0.32 \\ 
  Highest  & 0.14 & 0.16 & 0.21 & 0.23 & 0.21 \\ 
  Highest+ & 0.03 & 0.03 & 0.04 & 0.04 & 0.04 \\ 
  &  & & & & \\ 
  \textit{Education Level} &  & & & & \\ 
  Lowest & 0.00 & 0.00 & 0.00 & 0.00 & 0.00 \\ 
  Low  & 0.05 & 0.04 & 0.03 & 0.02 & 0.02 \\ 
  Medium  & 0.48 & 0.47 & 0.39 & 0.33 & 0.37 \\ 
  High & 0.18 & 0.20 & 0.19 & 0.18 & 0.19 \\ 
  Highest & 0.28 & 0.29 & 0.40 & 0.47 & 0.42 \\ 
  &  & & & & \\ 
  \textit{Sport Frequency} &  & & & & \\ 
  Never & 0.18 & 0.16 & 0.12 & 0.10 & 0.12 \\ 
  Rarely & 0.12 & 0.11 & 0.10 & 0.09 & 0.09 \\ 
  Monthly & 0.30 & 0.30 & 0.31 & 0.30 & 0.30 \\ 
  Weekly & 0.40 & 0.44 & 0.47 & 0.52 & 0.49 \\ 
  &  & & & & \\ 
  \midrule
   \cellcolor{white} \textit{Observations} &  & & & & \\ 
\midrule
  Total Share & 0.07 & 0.08 & 0.29 & 0.56 & 1.00 \\ 
  Total Observations & 4'690 & 5'827 & 19'970 & 39'429 & 69'916 \\ 
   \bottomrule
   \multicolumn{6}{l}{\footnotesize \textit{Note:} Means of variables displayed in all columns.}
\end{tabular}
\end{table}

\pagebreak

\section{Online Dating Platform}\label{App:A}

\subsection{Valid User Interactions}

In our analysis, we restrict ourselves to \textit{one-way} user interactions.
These interactions are always initiated by a visit from the sender, which is invisible to the recipient. The visit is then immediately followed by either a visible action from the sender, or possibly no further action at all. However, in both cases a visible reply of the recipient to this initial action by the sender is not permitted. In that sense, we retain only one-way interactions such that the sender was visibly or invisibly active, while the recipient stayed visibly passive. Hence, we do not allow for any visible reciprocal interaction between the sender and the recipient.

For instance, a sender visit followed by a sender message is a valid interaction. Also, two successive sender visits followed by a message is a valid interaction. A single sender visit is valid interaction, too. Further notice, that a sender visit followed by a recipient visit and afterwards a sender message is a valid interaction as well, as the sender has not seen the recipient's visit. However, a sender visit and sender like followed by a recipient visit and like back inducing a sender message is not a valid interaction anymore as the sender message has already been provoked by the recipient. Hence, we always restrict the interactions until the point a possible reciprocal interaction taking place.

\section{Additional Results}\label{App:D}

\subsection{Heterogeneous Effects}

\begin{table}[H]
\centering
\footnotesize
\caption{Wald Tests for Equality of Group Effects for Males and Females}\label{tab:wald}
\begin{tabular}{rrrrrr}
 \toprule
GATEs: Weekly vs. Never & \multicolumn{2}{c}{\cellcolor{white}Males} & & \multicolumn{2}{c}{\cellcolor{white}Females}\\
 \midrule
\cellcolor{white}\textit{Wald Test} & $\chi^2$ & $p$-Value & & $\chi^2$ & $p$-Value \\ 
   \midrule
\textit{Recipient Features} &  &  &  &  &  \\ 
   \midrule
  Age & 23.56 & 37.08 &  & 13.17 & 96.32 \\  
  Education Level & 14.03 & 0.72 & & 7.63 & 10.62  \\ 
  Income Level & 22.67  & 0.04 &  & 1.79 & 87.72  \\ 
  Sport Frequency & 9.15 & 2.74  &  & 3.43 & 32.94 \\ 
   \midrule
\textit{Sender Features} &  &  &  &  &  \\ 
   \midrule
Age & 32.18 & 7.45 &  & 24.18 & 50.90   \\  
  Education Level & 20.40 & 0.04 &  & 9.75 & 4.49 \\ 
  Income Level & 17.33 & 0.39 &  & 6.49 & 26.13  \\ 
  Sport Frequency & 6.55 & 8.76  &  & 4.41 & 0.24 \\ 
   \midrule
\textit{Shared Features} &  &  &  &  &  \\ 
   \midrule
Distance & 23.01 & 40.12 &  & 13.59 & 96.84  \\ 
   \bottomrule
   \multicolumn{6}{l}{\footnotesize \textit{Note:} Wald tests of Equality of the GATEs. $p$-Values in $\%$.}
   \end{tabular}
\end{table}

\pagebreak

\begin{ThreePartTable}
\begin{TableNotes}
\scriptsize
\item [] \hspace{-0.17cm} \footnotesize \textit{Note:} t-tests for Differences of the GATEs from the ATE. $\Delta$ in $\%$ points. $p$-Values in $\%$.
\end{TableNotes}
\scriptsize
\begin{longtable}{llrrrrlrrr}
\caption{Tests for Differences of GATEs to ATE for Males and Females}\label{tab:tests}\\
\toprule
GATEs: Weekly vs. Never & \multicolumn{4}{c}{\cellcolor{white}Males} & & \multicolumn{4}{c}{\cellcolor{white}Females}\\
 \midrule
\cellcolor{white}\textit{t-Test} & Group & $\Delta$ & SE & $p$-Value & & Group & $\Delta$ & SE & $p$-Value \\ 
 \midrule
\endfirsthead
\midrule
 \cellcolor{white}\textit{t-Test} & Group & $\Delta$ & SE & $p$-Value & & Group & $\Delta$ & SE & $p$-Value \\ 
 \midrule
 \endhead
 \midrule \multicolumn{10}{c}{\textit{continued on next page}} \\
\endfoot
\insertTableNotes
\endlastfoot
\textit{Recipient Features} & & & & & & & & &\\ 
   \midrule
  Age & 23.50 & -0.02 & 0.08 & 80.04 & & 21.00 & -0.09 & 0.27 & 75.02 \\ 
 & 30.00 & 0.02 & 0.07 & 83.72 & & 25.00 & -0.12 & 0.23 & 59.96\\ 
 & 31.50 & 0.06 & 0.07 & 40.97 & & 26.50 & 0.01 & 0.21 & 95.03\\ 
 & 33.00 & 0.04 & 0.07 & 56.74 & & 27.50 & -0.10 & 0.14 & 48.40\\ 
 & 34.50 & 0.05 & 0.06 & 45.19 & & 28.50 & -0.09 & 0.13 & 49.29\\ 
 & 35.50 & 0.03 & 0.09 & 76.92 & & 29.50 & -0.07 & 0.09 & 47.27\\ 
 & 36.50 & 0.06 & 0.08 & 49.74 & & 30.50 & -0.04 & 0.10 & 66.35\\ 
 & 37.50 & -0.01 & 0.04 & 83.13 & & 31.50 & 0.14 & 0.12 & 22.45\\ 
 & 39.00 & 0.04 & 0.05 & 40.91 & & 32.50 & -0.01 & 0.09 & 88.68 \\ 
 & 40.50 & 0.03 & 0.06 & 60.45 & & 33.50 & 0.13 & 0.10 & 17.17\\ 
 & 42.50 & -0.02 & 0.03 & 40.11 & & 34.50 & 0.06 & 0.06 & 36.87\\ 
 & 45.00 & -0.05 & 0.04 & 14.49 & & 35.50 & 0.07 & 0.09 & 45.64\\ 
 & 46.50 & -0.01 & 0.05 & 76.75 & & 36.50 & 0.07 & 0.06 & 26.29\\ 
 & 48.00 & -0.00 & 0.05 & 91.87 & & 37.50 & 0.12 & 0.07 & 9.96\\ 
 & 49.50 & 0.02 & 0.05 & 74.89 & & 38.50 & 0.03 & 0.08 & 69.35\\ 
 & 50.50 & 0.01 & 0.06 & 82.71 & & 40.00 & 0.01 & 0.08 & 93.72 \\ 
 & 51.50 & 0.00 & 0.06 & 94.83 & & 42.00 & -0.05 & 0.11 & 67.89 \\ 
 & 52.50 & -0.05 & 0.07 & 44.77 & & 43.50 & -0.09 & 0.13 & 48.02\\ 
 & 54.00 & 0.02 & 0.07 & 72.17 & & 45.00 & 0.07 & 0.11 & 51.04 \\ 
 & 55.50 & -0.06 & 0.07 & 41.80 & & 46.50 & 0.06 & 0.12 & 62.45 \\ 
 & 57.00 & -0.02 & 0.10 & 84.60 & & 48.00 & 0.03 & 0.12 & 80.54\\ 
 & 59.50 & -0.07 & 0.12 & 54.40 & & 49.50 & 0.02 & 0.14 & 89.29\\ 
 & 71.50 & -0.09 & 0.13 & 52.15 & & 51.00 & -0.02 & 0.14 & 88.52 \\ 
 &  &  &  &  & & 53.50 & -0.05 & 0.15 & 75.73\\
 &  &  &  &  & & 67.00 & -0.03 & 0.16 & 87.07\\
 & & & & & & & & &\\
 Education Level & Lowest & -0.15 & 0.11 & 17.30 & & Lowest & 0.11 & 0.29 & 70.80 \\ 
   & Low & -0.42 & 0.19 & 2.36 & & Low & -0.22 & 0.20 & 27.04 \\ 
   & Medium &-0.27 & 0.15 & 7.51 & & Medium & -0.10 & 0.07 & 19.95 \\ 
   & High & 0.03 & 0.02 & 15.84 & & High & 0.04 & 0.04 & 30.37 \\ 
   & Highest & 0.09 & 0.05 & 9.19 & & Highest & 0.09 & 0.10 & 35.32\\ 
    & & & & & & & & &\\
  Income Level & Lowest & -0.19 & 0.09 & 3.62 & & Lowest & -0.15 & 0.26 & 56.30\\ 
   & Low & -0.26 & 0.09 & 0.47 & & Low & -0.11 & 0.12 & 33.82 \\
   & Medium & -0.16 & 0.07 & 1.83 & & Medium & -0.00 & 0.04 & 95.63 \\
   & High & -0.03 & 0.02 & 13.33 & & High & 0.04 & 0.06 & 51.80 \\
   & Highest & 0.06 & 0.03 & 3.07 & & Highest & 0.08 & 0.10 & 41.01\\
   & Highest+ & 0.14 & 0.08 & 6.43 & & Highest+ & 0.09 & 0.13 & 51.46\\
     & & & & & & & & &\\
  Sport Frequency & Never &  -0.24 & 0.12 & 5.62 & & Never & -0.42 & 0.39 & 28.34\\ 
   & Rarely & -0.16  & 0.10 & 14.20 & & Rarely & -0.30 & 0.27 & 26.33\\ 
   & Monthly & -0.05 & 0.05 & 34.31 & & Monthly & -0.16 & 0.16 & 33.56 \\ 
   & Weekly & 0.03 & 0.02 & 26.99 & & Weekly & 0.09 & 0.09 & 30.43\\ 
   \midrule
\textit{Sender Features} & & & & & & & & &\\ 
   \midrule
Age & 22.50 & -0.03 & 0.09 & 76.98 & & 22.00 & -0.16 & 0.29 & 58.20 \\ 
 & 28.00 & 0.02 & 0.07 & 75.51 & & 27.00 & -0.08 & 0.22 & 71.84\\ 
 & 29.50 & 0.05 & 0.08 & 55.52 & & 28.50 & -0.14 & 0.17 & 40.21\\ 
 & 30.50 & 0.06 & 0.08 & 43.37 & & 29.50 & -0.14 & 0.15 & 36.84\\ 
 & 32.00 & 0.05 & 0.08 & 50.57 & & 30.50 & -0.01 & 0.11 & 90.90\\ 
 & 33.50 & 0.02 & 0.08 & 78.09 & & 31.50 & -0.04 & 0.12 & 72.55\\ 
 & 34.50 & 0.06 & 0.07 & 43.61 & & 32.50 & -0.03 & 0.09 & 73.61\\ 
 & 35.50 & 0.00 & 0.05 & 94.99 & & 33.50 & 0.03 & 0.08 & 66.48\\ 
 & 36.50 & 0.04 & 0.06 & 48.48 & & 34.50 & 0.02 & 0.07 & 82.77\\ 
 & 38.00 & 0.03 & 0.03 & 34.29 & & 35.50 & 0.09 & 0.07 & 17.47\\ 
 & 40.00 & -0.02 & 0.03 & 36.95 & & 36.50 & 0.16 & 0.08 & 4.37\\ 
 & 42.00 & -0.00 & 0.03 & 96.81 & & 37.50 & 0.06 & 0.05 & 28.04\\ 
 & 44.00 & -0.07 & 0.05 & 12.42 & & 38.50 & 0.06 & 0.06 & 34.36\\ 
 & 45.50 & -0.00 & 0.04 & 89.46 & & 40.00 & 0.10 & 0.06 & 8.29\\ 
 & 47.00 & -0.01 & 0.05 & 77.00 & & 41.50 & 0.01 & 0.07 & 87.26\\ 
 & 48.50 & -0.00 & 0.05 & 98.54 & & 43.00 & 0.05 & 0.07 & 52.76\\ 
 & 49.50 & 0.03 & 0.06 & 63.18 & & 44.50 & 0.00 & 0.08 & 95.30\\ 
 & 50.50 & 0.02 & 0.06 & 74.85 & & 46.00 & 0.04 & 0.09 & 69.01\\ 
 & 51.50 & 0.00 & 0.07 & 95.22 & & 47.50 & -0.01 & 0.11 & 95.12\\ 
 & 53.00 & -0.01 & 0.08 & 93.70 & & 48.50 & 0.07 & 0.11 & 48.48\\ 
 & 55.00 & -0.09 & 0.10 & 36.01 & & 50.00 & 0.01 & 0.12 & 91.68\\ 
 & 57.50 & -0.07 & 0.11 & 50.78 & & 51.50 & -0.10 & 0.13 & 46.68\\ 
 & 68.50 & -0.09 & 0.14 & 52.80 & & 53.00 & 0.02 & 0.12 & 88.61\\ 
 &  &  &  &  & & 55.00 & 0.01 & 0.13 & 91.37 \\
 &  &  &  &  & & 57.50 & 0.02 & 0.15 & 87.36\\
 &  &  &  &  & & 70.50 & -0.05 & 0.16 & 73.96\\
 & & & & & & & & &\\
  Education Level & Lowest & 0.09 & 0.10 & 36.18 & & Lowest & 0.04 & 0.15 & 75.74\\ 
   & Low & -0.39 & 0.16 & 1.36 & & Low & -0.22 & 0.15 & 13.95 \\ 
   & Medium & -0.10 & 0.06 & 8.91 & & Medium & -0.12 & 0.08 & 15.74\\ 
   & High & 0.00 & 0.02 & 91.15 & & High & -0.01 & 0.03 & 62.60 \\ 
   & Highest & 0.09 & 0.07 & 16.89 & & Highest & 0.10 & 0.06 & 8.18 \\ 
     & & & & & & & & &\\
  Income Level & Lowest & -0.17 & 0.08 & 3.20 & & Lowest & -0.14 & 0.19 & 46.40 \\ 
   & Low & -0.14 & 0.06 & 2.89 & & Low & -0.15 & 0.16 & 35.62 \\ 
   & Medium & -0.05 & 0.02 & 0.91 & & Medium & -0.04 & 0.07 & 58.00 \\ 
   & High & 0.02 & 0.02 & 22.64 & & High & -0.05 & 0.02 & 4.74 \\ 
   & Highest & 0.12 & 0.04 & 0.74 & & Highest & 0.07 & 0.06 & 28.41 \\ 
   & Highest+ & 0.15 & 0.06 & 1.22 & & Highest+ & 0.16 & 0.11 & 12.86 \\ 
     & & & & & & & & & \\
  Sport Frequency & Never & -0.10 & 0.06 & 10.27 & & Never & -0.24 & 0.09 & 1.17 \\ 
   & Rarely & -0.05 & 0.04 & 22.08 & & Rarely & -0.14 & 0.06 & 1.00 \\ 
   & Monthly & -0.00 & 0.01 & 64.62 & & Monthly & -0.05 & 0.02 & 2.98 \\ 
   & Weekly & 0.03 & 0.02 & 8.75 & & Weekly & 0.10 & 0.03 & 0.12 \\ 
   \midrule
\textit{Shared Features} & & & & & & & & &\\ 
   \midrule
Distance & 1.57 & 0.00 & 0.20 & 99.86 & & 1.45 & 0.03 & 0.11 & 79.53 \\
& 4.60 & 0.03   &    0.20 & 86.16  & & 4.26 & 0.01 & 0.11 & 89.61\\     
& 7.79 & 0.04    &   0.15 & 79.16  & & 7.09 & -0.02 & 0.11 & 88.02\\   
& 12.00 & 0.01    &   0.08 & 87.52   & & 10.38 & 0.05 & 0.07 &  48.18 \\ 
& 16.69 & 0.00   &  0.04 & 92.73  & & 14.24 & 0.03 & 0.06 & 58.33 \\
& 21.58 & 0.01   &  0.03 & 72.92  & & 18.09 & 0.06 & 0.05 &  19.38  \\
& 26.98 & 0.00   &  0.02 & 83.19  & & 22.23 & 0.01 & 0.04 &  85.48  \\
& 32.37 &-0.01   &   0.02 & 62.77  & & 26.85 & 0.02 & 0.03 &   63.64  \\
& 37.99 & 0.00   &  0.03 & 92.04  & & 31.26 & 0.02 & 0.03 &  55.16 \\
& 43.78 & 0.01    &   0.03 & 72.41   & & 35.62 & -0.04 & 0.03 &  15.40  \\
& 49.55 &-0.00 &  0.04 & 99.40   & & 40.15 & -0.03 & 0.03 & 35.87 \\
& 56.12 & 0.00   &   0.03 & 87.98   & & 44.60 & -0.03 & 0.04 &  50.06 \\ 
& 63.60 & -0.02   &    0.04 & 58.70  & & 49.10 & -0.04 & 0.03 & 24.65\\ 
& 71.36 & 0.01    &   0.04 & 72.91  & & 54.62 & -0.00 & 0.03 & 96.65 \\
& 79.30 & 0.00  &   0.04 & 98.34  & & 61.12 & -0.03 &  0.04 & 38.27\\
& 87.72 &-0.02   &    0.04 & 70.56  & & 67.78 & -0.02 & 0.03 & 55.05 \\
& 96.77 &-0.02   &    0.04 & 61.96  & & 74.76 & -0.02 & 0.04 & 62.73\\
& 108.00 &-0.04   &    0.04 & 39.33  & & 82.10 & 0.02 & 0.04 & 58.04\\
& 121.75 &-0.01   &    0.04 & 74.22  & & 89.77 & 0.01 & 0.05 & 84.55 \\
& 136.80 &-0.00  &   0.04 & 90.50   & & 97.78 & -0.00 & 0.04 & 99.96 \\
& 153.36 & 0.02    &   0.03 & 64.38   & & 107.76 & -0.01 & 0.05 & 86.38 \\
& 173.89 &-0.02   &   0.04 & 67.71   & & 120.60 & -0.03 & 0.05 & 60.48 \\
& 203.71 &-0.02   &   0.04 & 64.58   & & 134.74 & -0.00 & 0.05 & 99.07\\
&  &   &    &    & & 149.88 & 0.02 & 0.06 & 75.90\\
&  &   &   &    & & 170.12 & -0.01 & 0.05 & 92.23 \\
&  &   &    &    & & 202.07 & -0.02 & 0.06 & 78.49 \\
   \bottomrule
\end{longtable}
\end{ThreePartTable}

\begin{figure}[H]
    \centering
    \caption{Heterogeneous Effects of Sport Activity based on Income for Females}
    \label{fig:female_gates}
    \includegraphics[width=0.95\textwidth]{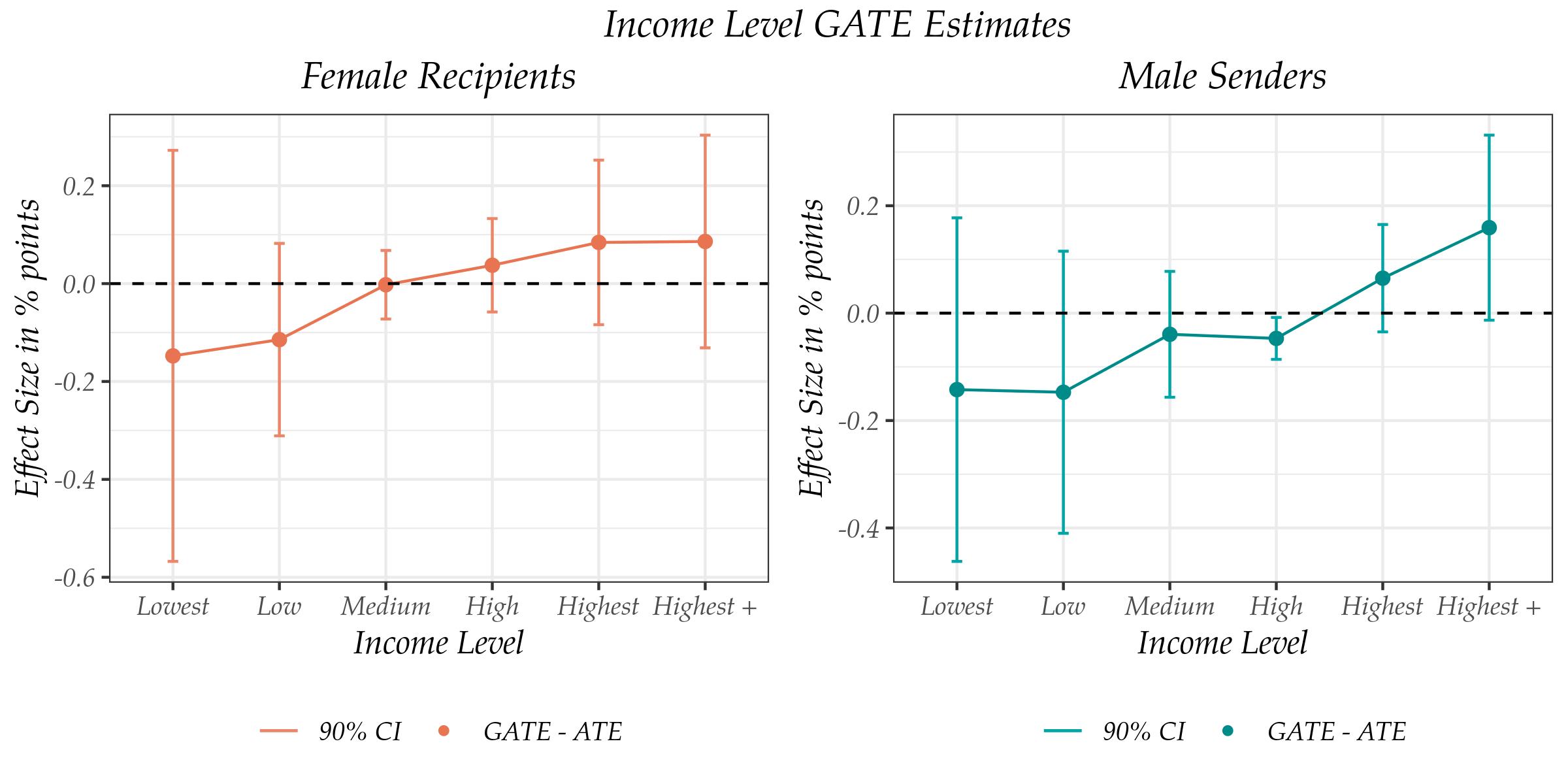}
    \caption*{\footnotesize \textit{Note:} Effects in $\%$ points as GATE deviations from the ATE (zero dotted line) with $90\%$ confidence intervals.}
\end{figure}

\begin{figure}[H]
    \centering
    \caption{Heterogeneous Effects of Sport Activity based on Sport for Males}
    \label{fig:male_gates_sport}
    \includegraphics[width=0.95\textwidth]{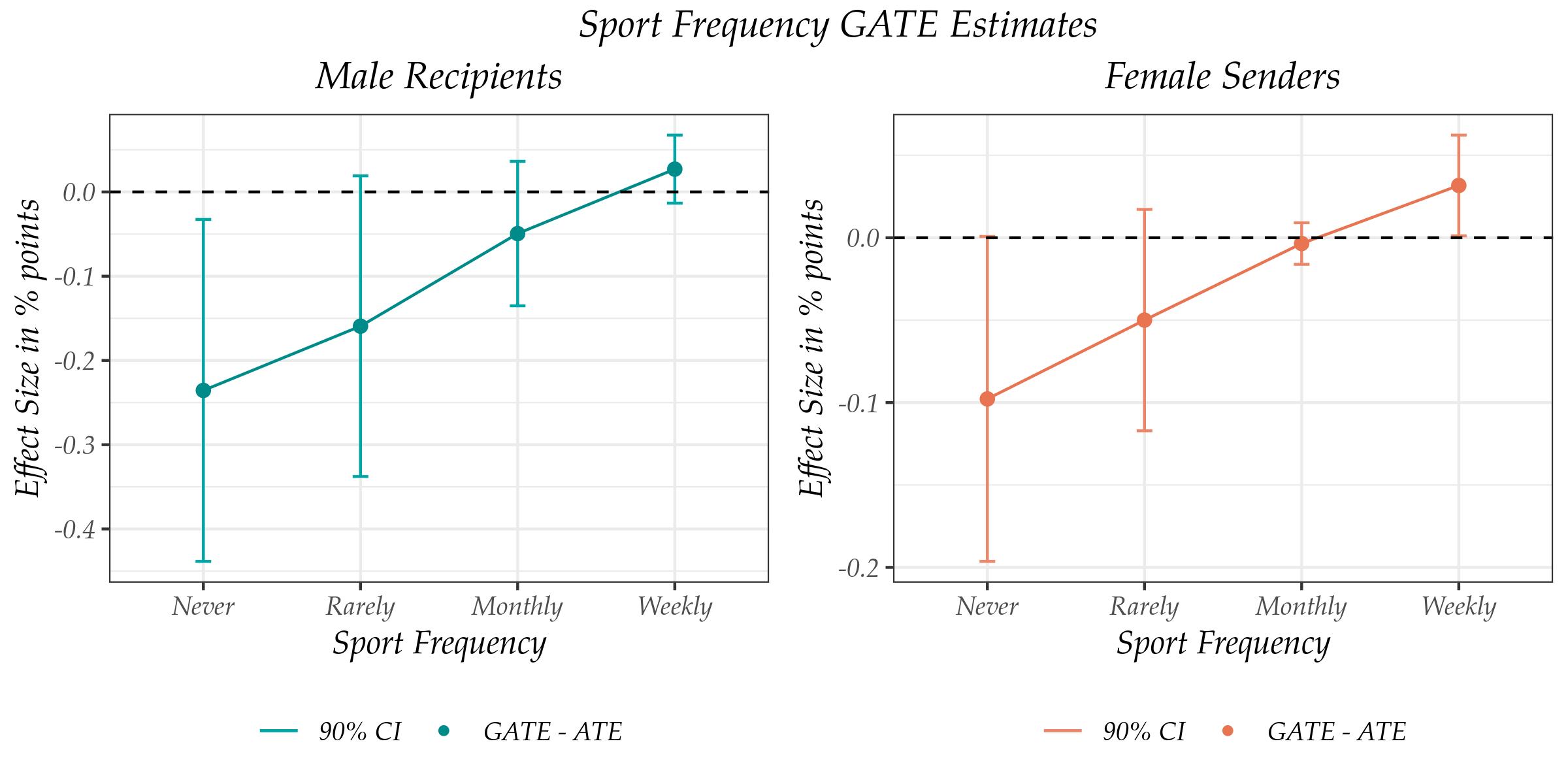}
    \caption*{\footnotesize \textit{Note:} Effects in $\%$ points as GATE deviations from the ATE (zero dotted line) with $90\%$ confidence intervals.}
\end{figure}

\begin{figure}[H]
    \centering
    \caption{Heterogeneous Effects of Sport Activity based on Age for Males}
    \label{fig:male_gates_age}
    \includegraphics[width=0.95\textwidth]{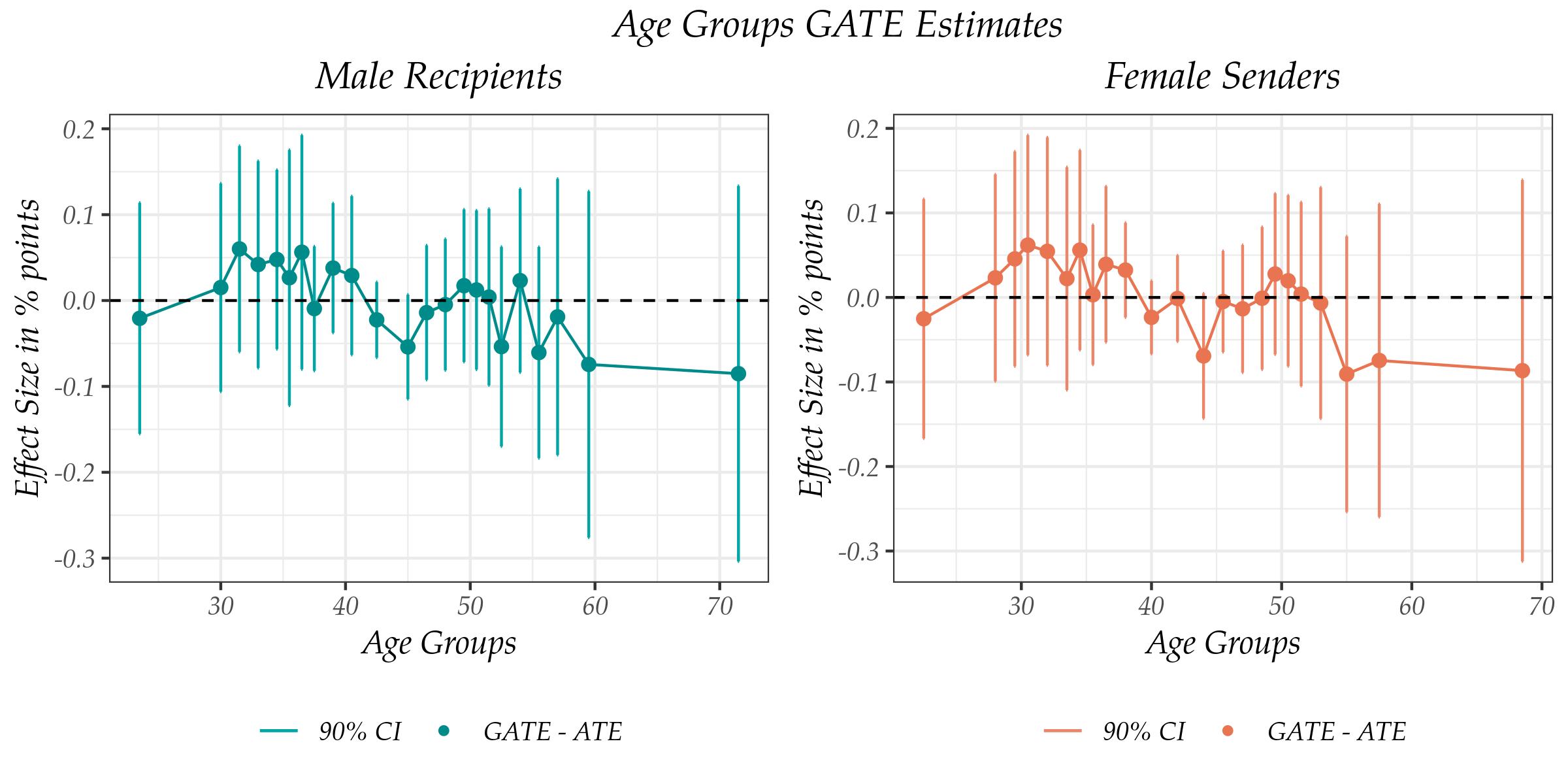}
    \caption*{\footnotesize \textit{Note:} Effects in $\%$ points as GATE deviations from the ATE (zero dotted line) with $90\%$ confidence intervals.}
\end{figure}

\begin{figure}[H]
    \centering
    \caption{Heterogeneous Effects of Sport Activity based on Age for Females}
    \label{fig:female_gates_age}
    \includegraphics[width=0.95\textwidth]{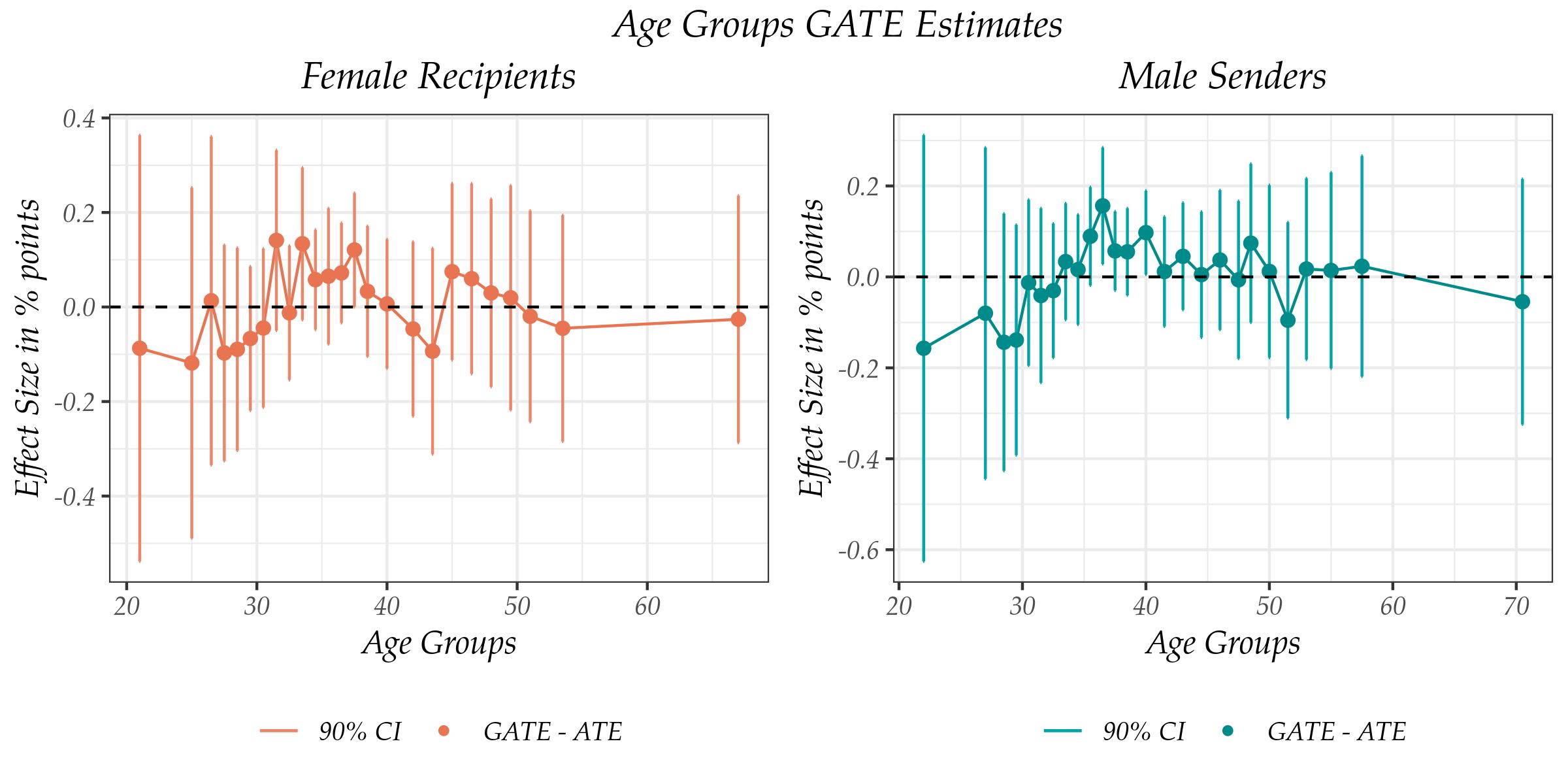}
    \caption*{\footnotesize \textit{Note:} Effects in $\%$ points as GATE deviations from the ATE (zero dotted line) with $90\%$ confidence intervals.}
\end{figure}

\begin{figure}[H]
    \centering
    \caption{Heterogeneous Effects of Sport Activity based on Education for Males}
    \label{fig:male_gates_edu}
    \includegraphics[width=0.95\textwidth]{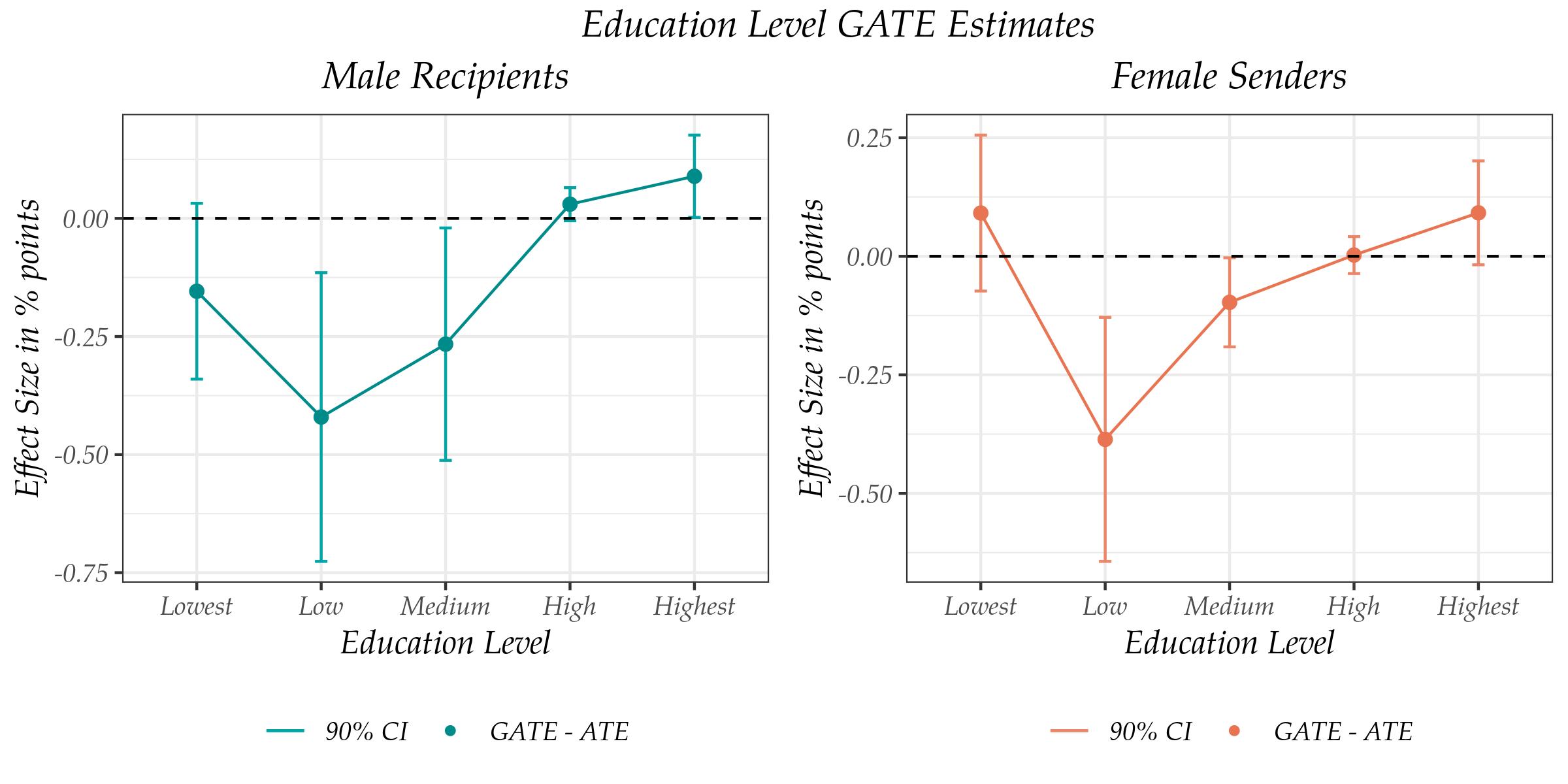}
    \caption*{\footnotesize \textit{Note:} Effects in $\%$ points as GATE deviations from the ATE (zero dotted line) with $90\%$ confidence intervals.}
\end{figure}

\begin{figure}[H]
    \centering
    \caption{Heterogeneous Effects of Sport Activity based on Education for Females}
    \label{fig:female_gates_edu}
    \includegraphics[width=0.95\textwidth]{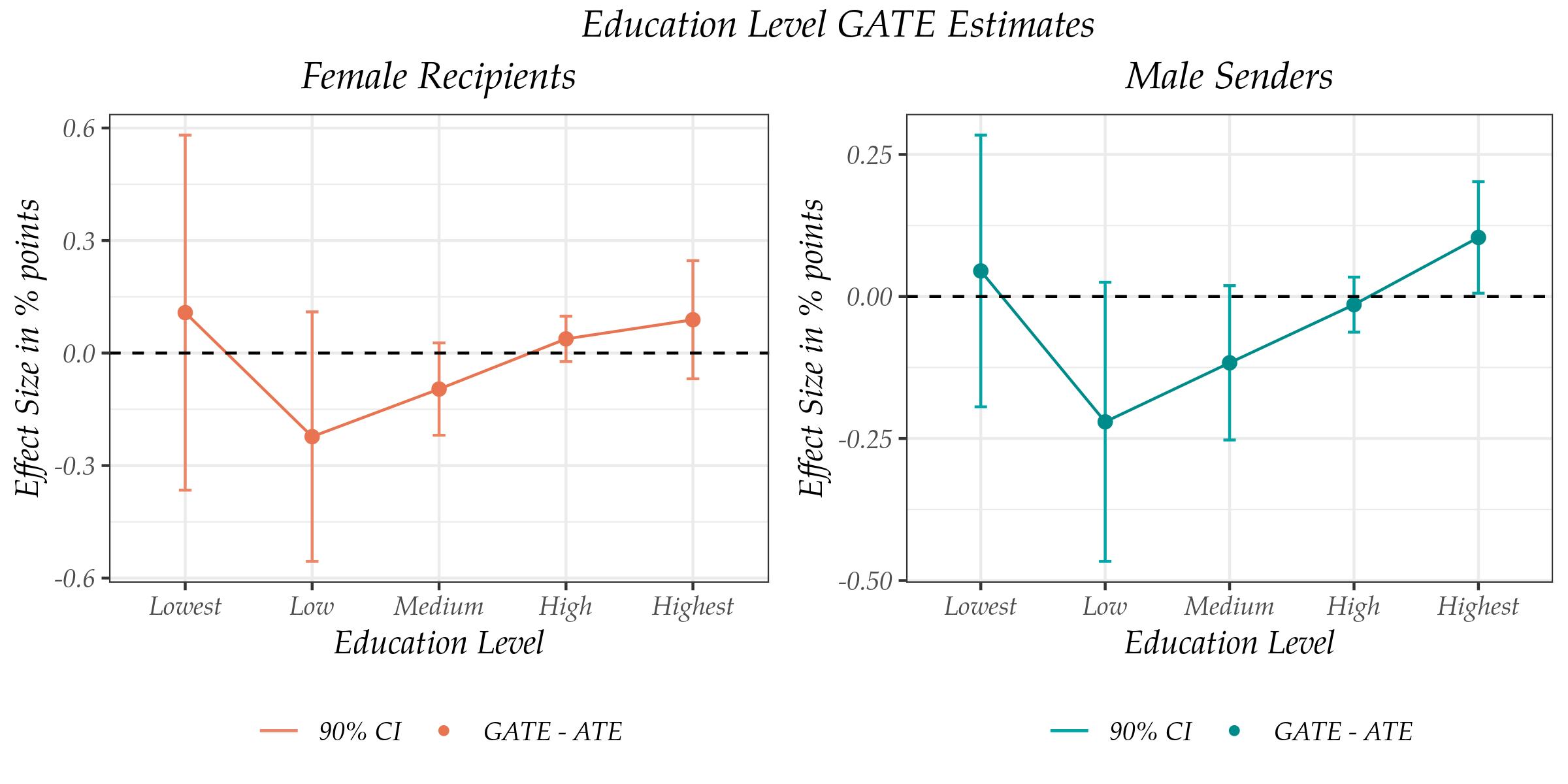}
    \caption*{\footnotesize \textit{Note:} Effects in $\%$ points as GATE deviations from the ATE (zero dotted line) with $90\%$ confidence intervals.}
\end{figure}

\begin{figure}[H]
    \centering
    \caption{Heterogeneous Effects of Sport Activity based on Distance}
    \label{fig:male_gates_distance}
    \includegraphics[width=0.95\textwidth]{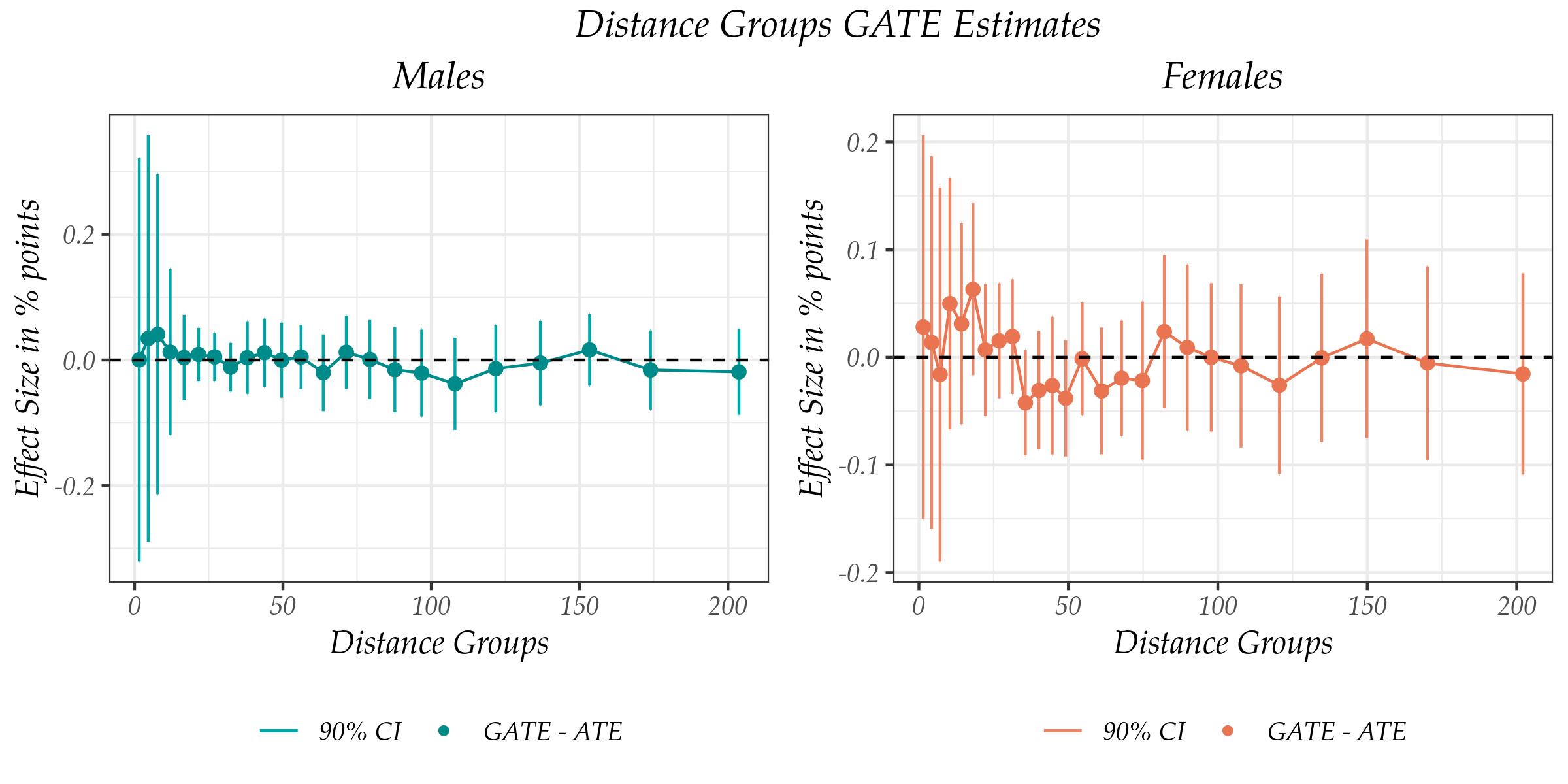}
    \caption*{\footnotesize \textit{Note:} Effects in $\%$ points as GATE deviations from the ATE (zero dotted line) with $90\%$ confidence intervals.}
\end{figure}

\subsection{Clustering Analysis}

\begin{table}[H]
\begin{adjustwidth}{-.5in}{-.5in}
\scriptsize
\centering
\caption{Descriptive Clusters of IATEs based on the $k$-means\plusplus{} Clustering}\label{tab:clusters}
\resizebox{1.1\textwidth}{!}{%
\begin{tabular}{llrrrrrlrrrrr}
 \toprule
Code & & \multicolumn{5}{c}{\cellcolor{white}Males} & & \multicolumn{5}{c}{\cellcolor{white}Females}\\
 \midrule
& \cellcolor{white}\textit{Clusters} & 1 & 2 & 3 & 4 & 5 & & 1 & 2 & 3 & 4 & 5 \\ 
     \midrule
& IATEs: Weekly vs. Never & 0.41 & 0.88 & 1.22 & 1.52 & 1.85 &  & -1.41 & -0.52 & 0.12 & 0.71 & 1.38 \\ 
   \midrule
& \textit{Recipient Features} &  &  &  &  &  &  &  &  &  &  &  \\ 
   \midrule
5 & Smoking Frequency & 1.30 & 0.85 & 0.32 & 0.10 & 0.03 &  & 0.50 & 0.47 & 0.45 & 0.37 & 0.29 \\ 
14 & Relevance of Sexuality & 0.38 & 0.44 & 0.47 & 0.49 & 0.52 &  & 0.21 & 0.27 & 0.31 & 0.32 & 0.31 \\ 
108 & TV in Leisure Time & 0.27 & 0.26 & 0.25 & 0.25 & 0.28 &  & 0.32 & 0.27 & 0.21 & 0.15 & 0.11 \\  
223 & Radio/TV at Home  & 0.62 & 0.62 & 0.60 & 0.59 & 0.60 &  & 0.73 & 0.68 & 0.65 & 0.64 & 0.67 \\ 
284 & Appearance Satisfaction & 0.18 & 0.19 & 0.20 & 0.21 & 0.22 &  & 0.11 & 0.16 & 0.18 & 0.20 & 0.21 \\
292 & Importance of Sexuality & 0.29 & 0.33 & 0.36 & 0.38 & 0.40 &  & 0.21 & 0.24 & 0.28 & 0.29 & 0.30 \\
303 & Comfortable Dining & 0.67 & 0.63 & 0.57 & 0.55 & 0.55 &  & 0.73 & 0.66 & 0.60 & 0.57 & 0.58 \\
324 & Wish Significant Other & 0.27 & 0.28 & 0.29 & 0.32 & 0.33 &  & 0.26 & 0.26 & 0.26 & 0.27 & 0.30 \\ 
     \midrule
& \textit{Sender Features}  &  &  &  &  &  &  &  &  &  &  &  \\ 
   \midrule
5 & Smoking Frequency & 0.69 & 0.52 & 0.36 & 0.31 & 0.27 &  & 0.48 & 0.50 & 0.49 & 0.42 & 0.32 \\ 
14 & Relevance of Sexuality & 0.24 & 0.28 & 0.29 & 0.33 & 0.40 &  & 0.35 & 0.40 & 0.44 & 0.45 & 0.46 \\ 
108 & TV in Leisure Time & 0.23 & 0.21 & 0.20 & 0.19 & 0.17 &  & 0.29 & 0.30 & 0.29 & 0.27 & 0.24 \\ 
223 & Radio/TV at Home & 0.65 & 0.61 & 0.60 & 0.58 & 0.55 &  & 0.67 & 0.66 & 0.64 & 0.63 & 0.61 \\ 
284 & Appearance Satisfaction & 0.18 & 0.17 & 0.17 & 0.19 & 0.22 &  & 0.12 & 0.16 & 0.17 & 0.19 & 0.23 \\
292 & Importance of Sexuality & 0.21 & 0.24 & 0.23 & 0.25 & 0.28 &  & 0.29 & 0.32 & 0.33 & 0.35 & 0.38 \\
303 & Comfortable Dining & 0.68 & 0.65 & 0.60 & 0.56 & 0.54 &  & 0.63 & 0.62 & 0.60 & 0.58 & 0.54 \\ 
324 & Wish Significant Other & 0.23 & 0.23 & 0.25 & 0.25 & 0.27 &  & 0.28 & 0.29 & 0.31 & 0.32 & 0.35 \\ 
   \midrule
& \textit{Observations} &  &  &  &  &  &  &  &  &  &  &  \\ 
   \midrule
& Share & 0.07 & 0.18 & 0.29 & 0.31 & 0.15 & & 0.07 & 0.20 & 0.29 & 0.29 & 0.15 \\ 
&  Total & 2288 & 6439 & 10153 & 10826 & 5252 & & 3753 & 11048 & 15896 & 15484 & 8047 \\ 
   \bottomrule
   \multicolumn{13}{l}{\scriptsize \textit{Note:} Means of clustered effects sorted in an increasing order, matched with selected user characteristics. Variable codes}\\
   \multicolumn{13}{l}{\scriptsize refer to the exact questions from the registration
   questionnaire presented in Table \ref{tab:questions}.}
   \end{tabular}%
   }
   \end{adjustwidth}
\end{table}

\pagebreak

\begin{landscape}
\pagestyle{empty}%

\section{Online Appendix}\label{App:online}

\subsection{Registration Questionnaire and Descriptive Statistics}

\begin{ThreePartTable}
\begin{TableNotes}
\scriptsize
\item [] \hspace{-0.17cm} \footnotesize \textit{Note:} First column lists a unique identifier for each variable. Second, third, fourth and fifth column report the corresponding mean, standard deviation, minimum and maximum values for the variable, respectively. Sixth column contains the specific questions from the registration questionnaire. Seventh column indicates the variable encoding: \textit{dummy} stands for a binary variable equal to 1 if the respective answer has been chosen (mutually inclusive); \textit{ordered} stands for a numeric value with a clear inherent ordering (both continuous and categorical), directly filled by the user (mutually exclusive); \textit{unordered} stands for a text value without an ordered structure (categorical), directly filled by the user (mutually exclusive). Last column lists the corresponding answers available in the registration questionnaire.
\end{TableNotes}
\begin{longtable}{|p{40pt}|>{\raggedleft\arraybackslash}p{30pt}|>{\raggedleft\arraybackslash}p{30pt}|>{\raggedleft\arraybackslash}p{30pt}|>{\raggedleft\arraybackslash}p{30pt}|p{150pt}|p{50pt}|p{150pt}|}
\caption{Summary of the Registration Questionnaire and the Descriptive Statistics}\label{tab:questions}\\
\hline
		\hline
		\textbf{Code} & \textbf{Mean} & \textbf{SD} & \textbf{Min} & \textbf{Max} & \textbf{Question} & \textbf{Coding} & \textbf{Answer} \\ 
		\hline
\endfirsthead
\hline
		Code & Mean & SD & Min & Max & Question & Coding & Answer \\ 
		\hline
 \endhead
 \hline \multicolumn{8}{c}{\textit{continued on next page}} \\
\endfoot
\insertTableNotes
\endlastfoot
		\hline
			No. 1 & 0.48 & 0.50 & 0.00 & 1.00 & Gender.                                                                                                                                                                                                                                                                                         & Dummy    &   Female                                                                                                                                                                                     \\ 
		\hline
			  No. 2 & 40.05 & 11.41 & 18.00 & 82.00 & Age (in years).                                                                                                                                                                                                                                                                                         & Ordered    &                                                                                                                                                                                        \\ 
		\hline
		\multirow{6}{*}{\setlength{\linewidth}{0cm}}No. 3 & \multirow{6}{*}{\setlength{\linewidth}{0cm}}3.47 & \multirow{6}{*}{\setlength{\linewidth}{0cm}}1.45 & \multirow{6}{*}{\setlength{\linewidth}{0cm}}1.00 & \multirow{6}{*}{\setlength{\linewidth}{0cm}}6.00  & \multirow{6}{*}{\setlength{\linewidth}{0cm}}Gross annual income (EUR).
		                                                                                                                                                                                                                                                   & Ordered      & Lowest income level ( 15,000 Euro)                                                                                                                                                     \\
\hline
\end{longtable}
\end{ThreePartTable}

\end{landscape}

\end{document}